\documentclass[11pt]{article}
\usepackage{bbm}

\usepackage{mathrsfs}
\usepackage{amsfonts}
\usepackage{amssymb}
\usepackage{amsmath,theorem,float,xypic}
\usepackage{color}

\newtheorem{theorem}{Theorem}[section]
\newtheorem{prop}[theorem]{Proposition}
\newtheorem{cor}[theorem]{Corollary}
\newtheorem{lemma}[theorem]{Lemma}
\newtheorem{definition}[theorem]{Definition}

\newtheorem{example}[theorem]{Example}

\newtheorem{defn}[theorem]{Definition}

\floatstyle{ruled}
\newfloat{algorithm}{t}{loa}
\floatname{algorithm}{Algorithm}
\newcommand{\Inp}[1]
  {\noindent\begin{tabular}{@{}p{1.8cm}@{}p{13.2cm}@{}}
   {\bf Input: }&#1 \end{tabular}}
\newcommand{\Outp}[1]
  {\noindent\begin{tabular}{@{}p{1.8cm}@{}p{13.2cm}@{}}
   {\bf Output: }&#1 \end{tabular}}

\def\imply{\Rightarrow}

\def\and{\cap}

\def\bref#1{(\ref{#1})}
\def\proof{{\noindent\em Proof:} }

\newcommand{\SPC}{\hspace*{15pt}}
\newcommand{\qedd}{\hspace*{\fill}$\Box$\medskip}
\newcommand{\bm}[1]{\mbox{\boldmath{$#1$}}}

\floatstyle{ruled}
\newfloat{algorithm}{t}{loa}
\floatname{algorithm}{Algorithm}

\def\st#1{{\em{#1}}}

\def\Y{{\mathbb{Y}}}
\def\U{{\mathbb{U}}}
\def\V{{\mathbb{V}}}
\def\L{{\mathbb{L}}}

\def\T{{\mathbb{T}}}

\def\AH{{\mathbb{A}}}
\def\IH{{\mathbb{I}}}
\def\CH{{\mathbb{C}}}

\def\C{{\mathcal C}}
\def\D{{\sigma}}

\def\G{{\mathbb G}}

\def\P{{\mathbb P}}
\def\Q{{\mathbb Q}}

\def\bu{{\mathbf{u}}}
\def\bv{{\mathbf{v}}}

\def\CC{{\mathbf{a}}}

\def\beps{{\bm\epsilon}}

\def\grem{\hbox{\rm grem}}

\def\sat{\hbox{\rm{sat}}}

\def\max{\hbox{\rm{max}}}

\def\gcd{\hbox{\rm{gcd}}}
\def\deg{\hbox{\rm{deg}}}

\def\ord{\hbox{\rm{ord}}}

\def\mod{\hbox{\rm{mod}}}

\def\ord{\hbox{\rm{ord}}}
\def\rk{\hbox{\rm{rk}}}

\def\coeff{\hbox{\rm{coeff}}}

\def\I{\mathcal{I}}

\def\LT{{\bf LT}}
\def\c{{\bf c}}
\def\b{{\bf b}}
\def\f{{\bf f}}
\def\g{{\bf g}}
\def\h{{\bf h}}

\def\a{{\bf a}}
\def\bv{{\bf v}}
\def\bu{{\bf u}}
\def\BU{{\bf u}}

\def\Z{{\mathbb{Z}}}
\def\Q{{\mathbb{Q}}}
\def\N{{\mathbb{N}}}

\def\CN{{\mathbb{C}}}
\def\A{{\mathbb{A}}}

\def\T{{\mathbb{T}}}
\def\Y{{\mathbb{Y}}}

\def\F{{\mathcal{F}}}
\def\E{{\mathcal{E}}}

\def\dtrdeg{\hbox{$\triangle$\rm{tr.deg}}}
\def\Jac{\hbox{\rm{Jac}}}

\def\Spec{\hbox{\rm{Spec}}}

\def\Hom{\hbox{\rm{Hom}}}
\def\image{\hbox{\rm{Im}}}
\def\Syz{\hbox{\rm{Syz}}}

\def\K{{\mathbb{K}}}

\def\grem{\hbox{\rm grem}}
\def\and{\cap}

\newcounter{bean}
\def\bl{\begin{list}{Step \arabic{bean}}{\usecounter{bean}}\labelwidth=34pt}
\def\el{\end{list}}

\def\deg{{\rm deg}}

\def\normalization1{{\rm normalization1}}
\def\normalization{{\rm normalization}}

\def\irrfactor1{{\rm irrfactor1}}
\def\irrfactor{{\rm irrfactor}}

\def\sat{{\rm sat}}

\def\rank{{\rm rk}}

\def\gb{Gr\"obner基}

\def\grem{\hbox{\rm{grem}}}

\def\st{\hbox{ {\rm{s.t.}} }}

\def\Zx{\Z[x]}
\def\Zxn{\Z[x]^n}

\def\fb{{\mathbbm{f}}}
\def\gb{{\mathbbm{g}}}

\begin{document}
\title{Toric Difference Variety}
\author{Xiao-Shan Gao, Zhang Huang, Jie Wang, Chun-Ming Yuan\\
 KLMM,  Academy of Mathematics and Systems Science\\
 Chinese Academy of Sciences, Beijing 100190, China}
\date{}
\maketitle

\begin{abstract}\noindent
In this paper, the concept of toric difference varieties is defined
and four equivalent descriptions for
toric difference varieties are presented in terms of difference
rational parametrization, difference coordinate rings,
toric difference ideals, and group actions by difference tori.
Connections between toric difference varieties
and affine $\N[x]$-semimodules are established by
proving the correspondence between the irreducible invariant difference subvarieties and the faces of the $\N[x]$-submodules and the orbit-face correspondence. Finally, an algorithm is given to decide whether a binomial difference ideal represented by a $\Zx$-lattice defines a toric difference variety.
%


\end{abstract}


\section{Introduction}
\label{sec-intro}

The theory of toric varieties has been extensively studied since its
foundation in the early 1970s by Demazure \cite{dema},  Miyake-Oda
\cite{oda}, Mumford et al. \cite{mum}, and Satake \cite{sata}, due
to its deep connections with polytopes, combinatorics, symplectic
geometry, topology, and its applications in physics, coding
theory, algebraic statistics, and hypergeometric functions
\cite{cox-2010,gelfand,oda-book}.
%

In this paper, we initiate the study of toric difference varieties and
expect that they will play similar roles in difference
algebraic geometry to their algebraic counterparts in
algebraic geometry.
Difference algebra and difference algebraic geometry
\cite{cohn, Hrushovski1, levin, wibmer} were founded by
Ritt \cite{ritt-dd1} and Cohn \cite{cohn}, who aimed to study
algebraic difference equations as algebraic geometry to polynomial
equations.

Similar to the algebraic case, a difference variety is called toric if
it is the Cohn closure of the values of a set of Laurent
difference monomials.
To be more precise, we introduce the notion of symbolic exponent.
For $p=\sum_{i=0}^s c_i x^i \in\Z[x]$ and $a$ in a difference field $k$ with
the difference operator $\sigma$,
denote $a^p = \prod_{i=0}^s (\sigma^i(a))^{c_i}.$
Then a Laurent difference monomial in the difference indeterminates
$\T=(t_1,\ldots,t_n)$ has the form $\T^{\bu}=\prod_{i=1}^n t_i^{u_i}$,
where $\bu=(u_1,\ldots,u_n) \in\Zx^n$.
For
\begin{equation}\label{eq-A0}
 U=\{\bu_1,\ldots, \bu_m\}, \hbox{ where }
 \bu_i \in \Z[x]^n,i=1,\ldots,m,
\end{equation}
define the following map
\begin{equation}
\phi_K\colon (K^*)^n\longrightarrow (K^*)^m, \T\mapsto\T^U=(\T^{\bu_1},\ldots,\T^{\bu_m}),
\end{equation}
where $K$ is any difference extension field of $k$ and $K^*=K\setminus\{0\}$.
Then, the toric difference variety $X_{U}$ defined by $U$
is the Cohn closure of the image of $\phi$.

A $\Zx$-lattice is a $\Zx$-submodule of $\Zx^n$, which plays the similar role
as lattice does in the study of  toric algebraic varieties.
A $\Zx$-lattice $L\subset\Zx^n$ is called toric if $g\bu\in L \imply \bu\in L$
for any $g\in\Zx\backslash\{0\}$ and $\bu\in \Zxn$.
We show that a difference variety $X\subset \AH^m$ is toric if and only if
the defining difference ideal for $X$ is
$\I_L=[\Y^{\bu}-\Y^{\bv}\mid \bu,\bv\in \N[x]^m \textrm{ with } \bu-\bv\in L]$
where $L$ is a toric $\Zx$-lattice and $\Y=(y_1,\ldots,y_m)$ is a set of
difference indeterminants.
%
%
An algorithm is given to decide whether
a $\Zx$-lattice is toric, and consequently,
to decide whether $\I_L$ defines a toric difference variety.

Similar to the algebraic case, a difference variety $X$ is toric
if and only if  $X$ contains a difference torus $T$ as a Cohn open subset
and with a difference algebraic group action of $T$ on $X$ extending the natural
group action of $T$ on itself.
Distinct from the algebraic case, a difference torus is not
necessarily isomorphic to $(\A^*)^m$, and this makes the definition of the difference torus more complicated.

Many properties of toric difference varieties
can be described using affine $\N[x]$-semimodules.
An affine $\N[x]$-semimodule $S$ generated by $U$ in \bref{eq-A0}
is $\{\sum_{i=1}^m g_i \bu_i\,|\, g_i\in \N[x] \}$.
It is shown that a difference variety $X$ is toric if and only
if  $X\simeq\Spec^{\sigma}(k[S])$,  where $S$ is an affine $\N[x]$-semimodule in $\Zx^n$ and $k[S]=$ $\{\sum_{\bu \in S}$
$\alpha_{\bu}\T^{\bu}\mid \alpha_{\bu}\in k, \alpha_{\bu}\ne0
\hbox{ for finitely many } \bu\}$.
Furthermore, there is a one-to-one correspondence between irreducible invariant subvarieties of a toric difference variety and faces of the corresponding affine $\N[x]$-semimodule.
A one-to-one correspondence between orbits of a toric difference variety and faces of the corresponding affine $\N[x]$-semimodule is also established.

Toric difference varieties connect difference Chow forms \cite{dd-chowform} and sparse
difference resultants \cite{dd-sres}. Precisely,
it is shown that the difference Chow form of $X_{U}$
is the difference sparse resultant of generic difference polynomials
with monomials ${\T^{\bu_1},\ldots,\T^{\bu_m}}$.
As a consequence, a Jacobi style order bound for a toric difference
variety $X_U$ is given.

The rest of this paper is organized as follows.
In section 2, preliminaries for difference algebra are introduced.
In section 3, the concept of difference toric variety  is defined and its coordinate ring is given
in terms of affine $\N[x]$-semimodules.
In section 4, the one-to-one correspondence between toric difference varieties
and toric difference ideals is given.
In section 5, a description of  toric difference varieties in terms of group action is given.
In section 6, deeper connections between toric difference varieties and  affine $\N[x]$-semimodules
are given.
In section 7, an order bound for a toric difference variety is given.
In section 8, an algorithm is given to decide whether a given $\Zx$-lattice is $\Zx$-saturated.
Conclusions are given in Section 9.

\section{Preliminaries}
%
We recall some basic notions from difference algebra. Standard references are \cite{cohn,levin,wibmer}. All rings in this paper will be assumed to be commutative and unital.

A {\em difference ring}, or {\em $\sigma$-ring} for short, is a ring $R$ together with a ring endomorphism $\sigma\colon R\rightarrow R$. If $R$ is a field, then we call it a {\em difference field}, or a {\em $\sigma$-field} for short.
A {\em morphism} between $\sigma$-rings $R$ and $S$ is a ring homomorphism $\psi\colon R\rightarrow S$ which preserves the difference operators.
In this paper, all $\sigma$-fields have characteristic $0$
and $k$ is a base $\sigma$-field.

A $k$-algebra $R$ is called a {\em $k$-$\sigma$-algebra} if the algebra structure map $k\rightarrow R$ is a morphism of $\sigma$-rings. A {\em morphism of $k$-$\sigma$-algebras} is a morphism of $k$-algebras which is also a morphism of $\sigma$-rings. A $k$-subalgebra of a $k$-$\sigma$-algebra is called a {\em $k$-$\sigma$-subalgebra} if it is stable under $\sigma$. If a $k$-$\sigma$-algebra is a $\sigma$-field, then it is called a {\em $\sigma$-field extension} of $k$. Let $R$ and $S$ be two $k$-$\sigma$-algebras. Then $R\otimes_k S$ is naturally a $k$-$\sigma$-algebra by defining $\sigma(r\otimes s)=\sigma(r)\otimes \sigma(s)$ for $r\in R$ and $s\in S$.

Let $k$ be a $\sigma$-field and $R$ a $k$-$\sigma$-algebra. For a subset $A$ of $R$, the smallest $k$-$\sigma$-subalgebra of $R$ containing $A$ is denoted by $k\{A\}$. If there exists a finite subset $A$ of $R$ such that $R=k\{A\}$, we say that $R$ is finitely $\sigma$-generated over $k$. If moreover $R$ is a $\sigma$-field, the smallest $k$-$\sigma$-subfield of $R$ containing $A$ is denoted by $k\langle A\rangle$.

Now we introduce the following useful notation. Let $x$ be an algebraic indeterminate and $p=\sum_{i=0}^s c_i x^i\in\Z[x]$. For $a$ in a $\sigma$-field, denote $a^p = \prod_{i=0}^s (\sigma^i(a))^{c_i}$ with $\sigma^0(a)=a$ and $a^0=1$. It is easy to check that $\forall p, q\in\Z[x], a^{p+q}=a^{p} a^{q}, a^{pq}= (a^{p})^{q}$.

Let $\Y=\{y_1,\ldots,y_m\}$ a set of $\sigma$-indeterminates over $k$. Then the {\em $\sigma$-polynomial ring} over $k$ in $\Y$ is the polynomial ring in the variables $\sigma^i(y_j)$
for $i\in\N$ and $j=1,\ldots,m$. It is denoted by
$k\{\Y\}=k\{y_1,\ldots,y_m\}$
and has a natural $k$-$\sigma$-algebra structure. A {\em $\sigma$-polynomial ideal}, or simply a {\em $\sigma$-ideal}, $I$ in $k\{\Y\}$ is an algebraic ideal which is closed under $\sigma$, i.e.\ $\sigma(I)\subset I$. If $I$ also has the property that $\sigma(a)\in I$ implies that $a\in I$, it is called a {\em reflexive $\sigma$-ideal}. A {\em $\sigma$-prime} ideal is a reflexive $\sigma$-ideal which is prime as an algebraic ideal. A $\sigma$-ideal $I$ is called {\em perfect} if for any $g\in\N[x]\setminus\{0\}$ and $a\in k\{\Y\}$, $a^g\in I$ implies $a\in I$. It is easy to prove that every $\sigma$-prime ideal is perfect. If $S$ is a finite set of $\sigma$-polynomials in $k\{\Y\}$, we use $(S)$, $[S]$, and $\{S\}$ to denote the algebraic ideal, the $\sigma$-ideal, and the perfect $\sigma$-ideal generated by $S$ respectively.

For $\bu=(u_1,\ldots,u_m)\in \Z[x]^m$, $\Y^{\bu}=\prod_{i=1}^m y_i^{u_i}$ is called a Laurent {\em $\sigma$-monomial} and $\bu$ is called its {\em support}.
A Laurent {\em $\sigma$-polynomial} in $\Y$ is a linear combination of
Laurent $\sigma$-monomials and $k\{\Y^{\pm}\}$ denotes the set of all  Laurent $\sigma$-polynomials, which is a $k$-$\sigma$-algebra.

Let $k$ be a $\sigma$-field. We denote the category of $\sigma$-field extensions of $k$ by $\mathscr{E}_k$ and the category of $K^m$ by $\mathscr{E}_k^m$ where $K \in\mathscr{E}_k$. Let $F\subset k\{\Y\}$ be a set of $\sigma$-polynomials. For any $K \in\mathscr{E}_k$, define the solutions of $F$ in $K$ to be
$$\mathbb{V}_K(F):=\{a\in K^m\mid f(a)=0 \textrm{ for all } f\in F\}.$$
Note that $K\rightsquigarrow \V_K(F)$ is naturally a functor from the category of $\sigma$-field extension of $k$ to the category of sets. Denote this functor by $\V(F)$.
\begin{definition}
Let $k$ be a $\sigma$-field. An {\em (affine) difference variety} or $\sigma$-variety over $k$ is a functor $X$ from the category of $\sigma$-field extension of $k$ to the category of sets which is of the form $\mathbb{V}(F)$ for some subset $F$ of $k\{\Y\}$. In this situation, we say that $X$ is the (affine) $\sigma$-variety defined by $F$.
If there is no confusion, we will omit the word ``affine'' for short.
\end{definition}

The functor $\mathbb{A}_k^m$ given by $\mathbb{A}_k^m(K)=K^m$ for $K\in \mathscr{E}_k$ is called the $\sigma$-affine ($n$-)space over $k$. If the base field $k$ is specified, we often omit the subscript $k$.


Let $X$ be a subset of $\mathbb{A}_k^m$. Then
\[\mathbb{I}(X):=\{f\in k\{\Y\}\mid f(a)=0\textrm{ for all }a\in X(K) \textrm{ and all }K\in \mathscr{E}_k\}\]
is called the {\em vanishing ideal} of $X$.
It is well known that $\sigma$-subvarieties of $\mathbb{A}_k^m$ are in a one-to-one correspondence with perfect $\sigma$-ideals of $k\{\Y\}$ and we have $\mathbb{I}(\V(F))=\{F\}$ for $F\subset k\{\Y\}$.

\begin{definition}
Let $X$ be a $\sigma$-subvariety of $\mathbb{A}_k^m$. Then the $k$-$\sigma$-algebra
\[k\{X\}:=k\{\Y\}/\mathbb{I}(X)\]
is called the {\em $\sigma$-coordinate ring} of $X$.
\end{definition}

A $k$-$\sigma$-algebra isomorphic to some $k\{\Y\}/\mathbb{I}(X)$ is called an {\em affine $k$-$\sigma$-algebra}. By definition, $k\{X\}$ is an affine $k$-$\sigma$-algebra.
Similar to affine algebraic varieties,
the category of affine $k$-$\sigma$-varieties is antiequivalent to the category of affine $k$-$\sigma$-algebras \cite{wibmer}.
The following lemma is from \cite[p.27]{wibmer}.
\begin{lemma}\label{pdag-lemma1}
Let $X$ be a $k$-$\sigma$-variety. Then for any $K\in \mathscr{E}_k$, there is a natural bijection between $X(K)$ and the set of $k$-$\sigma$-algebra homomorphisms from $k\{X\}$ to $K$. Indeed,
$$X\simeq \Hom(k\{X\},\A^1)$$
as functors.
\end{lemma}

Suppose that $k\{X\}$ is an affine $k$-$\sigma$-algebra. Let $\Spec^{\sigma}(k\{X\})$ be the set of all $\sigma$-prime ideals of $k\{X\}$. Let $F\subseteq k\{X\}$. Set
$$\mathcal{V}(F):=\{\mathfrak{p}\in \Spec^{\sigma}(k\{X\})\mid F\subset \mathfrak{p}\}\subset \Spec^{\sigma}(k\{X\}).$$
It can be checked that $\Spec^{\sigma}(k\{X\})$ is a topological space with closed sets of the form $\mathcal{V}(F)$. Then the {\em topological space} of $X$ is $\Spec^{\sigma}(k\{X\})$ equipped with the above Cohn topology.

Let $k$ be a $\sigma$-field and $F\subset k\{\Y\}$. Let $K,L\in\mathscr{E}_k$. Two solutions $a\in\V_K(F)$ and $b\in\V_L(F)$ are called {\em equivalent} if there exists a $k$-$\sigma$-isomorphism between $k\langle a\rangle$ and $k\langle b\rangle$ which maps $a$ to $b$. Obviously this defines an equivalence relation.
The following theorem gives a relationship between equivalence classes of solutions of $I$ and $\sigma$-prime ideals containing $I$. See \cite[p.31]{wibmer}.
\begin{theorem}\label{pdag-thm}
Let $X$ be a $k$-$\sigma$-variety. There is a natural bijection between the set of equivalence classes of solutions of $\mathbb{I}(X)$ and $\Spec^{\sigma}(k\{X\})$.
\end{theorem}

We shall not strictly distinguish between a $\sigma$-variety and its topological space. In other words, we use $X$ to mean the $\sigma$-variety or its topological space.

\section{Affine toric $\sigma$-varieties}
In this section, we will define affine toric $\sigma$-varieties and
give a description of their coordinate rings in terms of affine $\N[x]$-semimodules.

Let $k$ be a $\sigma$-field. Let $(\A^*)^n$ be the functor from $\mathscr{E}_k$ to $\mathscr{E}_k^n$ satisfying $(\A^*)^n(K)=(K^*)^n$ where $K\in \mathscr{E}_k$ and $K^*=K\backslash \{0\}$. In the rest of this section, always assume \begin{equation}\label{eq-U}
U=\{\bu_1,\ldots,\bu_m\}\subseteq \Z[x]^n
 \hbox{ and } \T=\{ t_1,\ldots,t_n\}
\end{equation}
 a set of $\sigma$-indeterminates. We define the following map
\begin{equation}\label{tvsm-equ1}
\phi\colon(\A^*)^n \longrightarrow (\A^*)^m,  \T \mapsto\T^{U} = (\T^{\bu_1},  \ldots, \T^{\bu_m}).
\end{equation}

Define the functor $T_{U}^*$ from $\mathscr{E}_k$ to $\mathscr{E}_k^m$ with $T_{U}^*(K)=\image(\phi_K)$ for each $K\in \mathscr{E}_k$ which is called the {\em quasi $\sigma$-torus} defined by $U$.
\begin{definition}
An affine $\sigma$-variety over the $\sigma$-field $k$ is called {\em toric} if it is the Cohn closure of a quasi $\sigma$-torus
$T_{U}^*\subseteq \A^m$ in $\A^m$.
Precisely, let
\begin{equation}\label{eq-atv0}
\I_U:=\{f\in k\{\Y\}=k\{y_1,\ldots,y_m\}\mid f(\T^{\bu_1}, \ldots, \T^{\bu_m})=0\}.
\end{equation}
Then the (affine) toric $\sigma$-variety defined by $U$ is $X_{U}=\V(\I_{U})$.
The matrix $\overline{U}=[\bu_1,\ldots,\bu_m]$  with $\bu_i$ as the $i$-th column is called the matrix representation for $X_U$.
\end{definition}


\begin{lemma}\label{lm-tv10}
$X_U$ defined above is an irreducible $\sigma$-variety of $\D$-dimension $\rank(\overline{U})$, where $\overline{U}$ is the matrix representation of $X_U$.
\end{lemma}
\proof
It is clear that $\T^U$ in \bref{tvsm-equ1} is a generic zero of $\I_U$ in \bref{eq-atv0}. Then $\I_U$ is a $\D$-prime $\sigma$-ideal. By Theorem 3.20 of \cite{dd-sres}, $\I_U$ is of $\D$-dimension $\dtrdeg\, k\langle\T^U\rangle/k =\rk(\overline{U})$.
\qedd

Let $\T^{\pm} = \{ t_1, \ldots, t_n, t_1^{-1}, \ldots, t_n^{-1}\}$.
Let $\I_{U,\T^{\pm}}=[y_1-\T^{\bu_1},\ldots,y_m-\T^{\bu_m}]$ be the $\sigma$-ideal generated by $y_i-\T^{\bu_i},i=1,\ldots,m$ in $k\{\Y,\T^{\pm}\}$. Then it is easy to check
\begin{equation}\label{eq-tv01}
\I_U=\I_{U,\T^{\pm}}\cap k\{\Y\}.
\end{equation}
Alternatively, let $z$ be a new $\sigma$-indeterminate and
$\I_{U,\T}=[\T^{\bu_1^+}y_1-\T^{\bu_1^-},\ldots,\T^{\bu_m^+}y_m-\T^{\bu_m^-}, \prod_{i=1}^nt_i z-1]$
be a $\sigma$-ideal in $k\{\Y,\T,z\}$, where $\bu_i^+,\bu_i^-\in\N[x]^n$ are the
postive and negative parts of $\bu_i=\bu_i^+-\bu_i^-$, respectively, $i=1,\ldots,m$. Then
\begin{equation}\label{eq-tv02}
\I_U=\I_{U,\T}\cap k\{\Y\}.
\end{equation}

Equation \bref{eq-tv02} can be used to compute a characteristic set \cite{gao-dcs}
for $\I_U$ as shown in the following example.
\begin{example}\label{ex-tv1}
Let $M=\left[\begin{array}{llll}
 2       & x-1      & 0      & 0   \\
 0       & 0          & 2      & x-1       \\
\end{array}\right]$
and $U$ the set of column vectors of $M$. Let
$\I_{1}=[y_1-t_1^2, t_1y_2-t_1^{x},y_3-t_2^2,t_2y_4-t_2^{x}, t_1t_2z-1]$.
By \bref{eq-tv02}, $\I_U = \I_1\cap k\{y_1,y_2,y_3,y_4\}$.
With the characteristic set method \cite{gao-dcs}, under the variable order $y_2 < y_4<y_1 < y_3< t_1 < t_2<z$, a characteristic set of $\I_1$ is $y_1y_2^{2}-y_1^x, y_{3}y_{4}^{2}-y_{3}^x,
y_1-t_1^2,t_1y_2-t_1^{x},y_3-t_2^2, t_2y_4-t_2^{x}, t_1t_2z-1$.
Then $\I_U =\I_1\cap k\{y_1,y_2,y_3,y_4\} =[y_1y_2^{2}-y_1^x,y_{3}y_{4}^{2}-y_{3}^x].$
\end{example}

The following example shows that some $y_i$ might not appear effectively in $\I_U$.
\begin{example}\label{ex-tv2}
Let $U= \{[1,1]^\tau,[x,x]^\tau,[0,1]^\tau\}$.
By \bref{eq-tv02}, $\I_U=[y_1-t_1t_2,y_2-t_1^{x}t_2^x,y_3-t_2,t_1t_2z-1]\cap k\{y_1,y_2,y_3\}=[y_1^x-y_2]$
and $y_3$ does not appear in $\I_U$.
\end{example}

Next, we will give a description for the
coordinate ring of a toric $\sigma$-variety
in terms of affine $\mathbb{N}[x]$-semimodules.
$S\subseteq \mathbb{Z}[x]^n$ is called an {\em $\mathbb{N}[x]$-semimodule} if it satisfies (i) $\a+\b\in S, \forall \a,\b\in S$; (ii) $g\a\in S, \forall g\in \mathbb{N}[x], \forall \a\in S$. Moreover, if there exists a finite subset $U=\{\bu_1,\dots,\bu_m\}\subset \mathbb{Z}[x]^n$ such that $S=\mathbb{N}[x](U)=\{\sum_{i=1}^m g_i\bu_i\mid g_i\in \mathbb{N}[x]\}$, $S$ is called an {\em affine $\mathbb{N}[x]$-semimodule}.
A map $\phi\colon S\rightarrow S'$ between two $\mathbb{N}[x]$-semimodules is an {\em $\mathbb{N}[x]$-semimodule morphism} if $\phi(\a+\b)=\phi(\a)+\phi(\b),\phi(g\a)=g\phi(\a)$ for all $\a,\b\in S,g\in \mathbb{N}[x]$ and $\phi(0)=0$.

Let $k$ be a $\sigma$-field. For every affine $\mathbb{N}[x]$-semimodule $S$, we associate it with the following {\em $\mathbb{N}[x]$-semimodule algebra} $k[S]$ which is the vector space over $k$ with $S$ as a basis and multiplication induced by the addition of $S$. More concretely,
\[k[S]:=\bigoplus_{\bu\in S}k\T^{\bu}=\{\sum_{\bu\in S}c_{\bu}\T^{\bu}\mid c_{\bu}\in k \textrm{ and } c_{\bu}=0 \textrm{ for all but finitely many } \bu\}\]
with multiplication induced by
$\T^{\bu}\cdot\T^{\bv}=\T^{\bu+\bv},\forall \bu,\bv\in S$.
Make $k[S]$ to be a $k$-$\sigma$-algebra by defining
$\sigma(\T^{\bu})=\T^{x\bu},\forall \bu\in S.$

If $S=\N[x](U)=\N[x](\{\bu_1,\ldots,\bu_m)\}$, then $k[S]=k\{\T^{\bu_1},\ldots,\T^{\bu_m}\}$. Therefore, $k[S]$ is a finitely $\sigma$-generated $k$-$\sigma$-algebra. When an embedding $S\rightarrow \Z[x]^n$ is given, it induces an embedding $k[S]\rightarrow k[\Z[x]^n] \simeq k\{t_1^{\pm 1},\ldots,t_n^{\pm 1}\}=k\{\T^{\pm}\}$. So $k[S]$ is a $k$-$\sigma$-subalgebra of $k\{\T^{\pm}\}$ generated by finitely many Laurent $\sigma$-monomials and it follows that $k[S]$ is a $\sigma$-domain. We will see that $k[S]$ is actually the $\sigma$-coordinate ring of a toric $\sigma$-variety.
\begin{theorem}\label{astv-thm1}
Let $X$ be an affine $\sigma$-variety. Then $X$ is a toric $\sigma$-variety if and only if there exists an affine $\N[x]$-semimodule $S$ such that $X\simeq \Spec^{\sigma}(k[S])$. Equivalently, the $\sigma$-coordinate ring of $X$ is $k[S]$.
\end{theorem}
\proof Let $X=X_U$ be a toric $\sigma$-variety defined
by $U$ in \bref{eq-U} and $\I_U$ defined in \bref{eq-atv0}.
Let $S=\N[x](U)$ be the affine $\N[x]$-semimodule
generated by $U$.
Define the following morphism of $\sigma$-rings
$$
\theta: k\{\Y\}\longrightarrow k[S], \hbox{ where }
\theta(y_{i})= \T^{\bu_i},i=1,\ldots,m.
$$
The map $\theta$ is surjective by the definition of $k[S]$. If
$f\in \ker(\theta)$,  then
 $f(\T^{\bu_1}, \ldots, \T^{\bu_i})=0$, which is
equivalent to $f\in \I_U$. Then, $\ker(\theta) =\I_U$
and $k\{\Y\}/{\I_U} \simeq k[S]$. Therefore
$X\simeq\Spec^{\sigma}(k\{\Y\}/\I_U)=\Spec^{\sigma}(k[S])$.
Conversely, if $X\simeq\Spec^{\sigma}(k[S])$,  where $S \subseteq \Z[x]^n$
is an affine $\N[x]$-semimodule, and $S=\N[x](\{\bu_{1}, \ldots,$
$\bu_{m}\})$ for $\bu_{i}\in S$.
Let $X_U$ be the toric $\sigma$-variety defined by
$U=\{\bu_{1}, \ldots, \bu_{m}\}$. Then as we just proved,
the coordinate ring of $X$ is isomorphic to $k[S]$. Then
$X \simeq X_{U}$.\qedd

We further have
\begin{prop}\label{prop-oo}
Let $S=\N[x](\{\bu_1,\ldots,\bu_m\})\subset\Zx^n$ be an affine $\mathbb{N}[x]$-semimodule and let $X=\Spec^{\sigma}(k[S])$ be the toric $\sigma$-variety associated with $S$.
Then there is a one-to-one correspondence between $X(K)$ and $\Hom(S,K)$, $\forall K\in \mathscr{E}_k$. Equivalently, $X\simeq \Hom(S, \A^1)$.
\end{prop}
\proof
By Lemma \ref{pdag-lemma1}, an element of $X(K)$ is given by a $k$-$\sigma$-algebra homomorphism $f\colon k[S]\rightarrow K$, where $K\in \mathscr{E}_k$. This corresponds to such a morphism $\varphi\colon S\rightarrow K$ satisfying $\varphi(\sum_ig_i\bu_i)=\prod_i\varphi(\bu_i)^{g_i},\forall \bu_i\in S,\forall g_i\in \mathbb{N}[x]$ such that $f(\T^{\bu_i})=\varphi(\bu_i)$.
\qedd

In the rest of this paper, we will identity elements of $X(K)$ with
morphisms from $S$ to $K$ and use $\phi,\psi,\gamma$ to denote these elements.

\section{Toric $\sigma$-ideal}\label{sec-tv2}
In this section,  we will show that $\sigma$-toric varieties are
defined exactly by toric $\sigma$-ideals. We first define the concept of
$\Zx$-lattice which is introduced in \cite{gao-dbi}.

A $\Z[x]$-module which can be embedded into $\Z[x]^m$ for some $m$ is called a {\em $\Z[x]$-lattice}. Since $\Z[x]^m$ is Noetherian as a $\Z[x]$-module, we see that any $\Z[x]$-lattice is finitely generated. Let $L$ be generated by $\fb=\{\f_1,\ldots,\f_s\}\subset\Zx^m$, which is denoted as $L=(\fb)_{\Zx}$. Then the matrix with $\f_i$ as the $i$-th column
is called a {\em matrix representation} of $L$.
%
%
Define the {\em rank} of $L$ $\rank(L)$ to be the rank of its representing matrix.
Note $L$ may not be a free $\Z[x]$-module, thus the number of minimal generators of $L$ could be larger than its rank.

A $\Z[x]$-lattice $L\subseteq \Z[x]^{m}$ is called {\em toric} if it is $\Z[x]$-saturated, that is for any nonzero $g\in\Z[x]$ and $\bu \in\Z[x]^{m}$, $g\bu \in L$ implies $\bu \in L$.

\begin{definition}
Associated with a $\Z[x]$-lattice $L\subseteq \Z[x]^m$, we defined a binomial $\sigma$-ideal $\I_L\subseteq k\{\Y\}=k\{y_1,\ldots,y_m\}$
\[\I_L:=[\Y^{\bu^+}-\Y^{\bu^-}\mid\bu\in L]=[\Y^{\bu}-\Y^{\bv}\mid \bu,\bv\in \N[x]^m \textrm{ with } \bu-\bv\in L],\]
where $\bu^{+}, \bu^{-}\in \N[x]^m$ are the
positive part and the negative part of $\bu=\bu^{+}-\bu^{-}$, respectively.
If $L$ is toric, then the corresponding $\Z[x]$-lattice ideal $\I_L$ is called a {\em toric $\sigma$-ideal}.
\end{definition}


$\I_L$ has the following properties.
\begin{itemize}
\item
Since a toric $\Zx$-lattice is both $\Z$-saturated and $x$-saturated,
by Corollary 6.22 in \cite{gao-dbi}, $\I_L$ is a $\sigma$-prime ideal of $\sigma$-dimension $m-\rank(L)$.

\item
By Theorem 6.19 in \cite{gao-dbi}, toric $\sigma$-ideals $I_L$ in $k\{\Y\}$ are in a one-to-one correspondence with toric $\Z[x]$-lattices $L$ in $\Z[x]^m$, that is,
    $L=\{\bu-\bv\,|\, \Y^\bu - \Y^\bv \in \I_L \}$ .
    $L$ is called the {\em support lattice} of $\I_L$.
\end{itemize}

In the rest of this section, we will prove the following result which
 can be  deduced from Lemmas \ref{lm-tvi1} and \ref{lm-tvi4}.
\begin{theorem}\label{th-tvi}
A $\sigma$-variety $X$ is toric if and only if $\IH(X)$ is
a toric $\sigma$-ideal.
\end{theorem}

\begin{lemma}\label{lm-tvi1}
Let $X_U$ be the toric $\sigma$-variety defined in \bref{eq-atv0}.
Then $\I_U=\IH(X_{U})$ is a toric
$\sigma$-ideal whose support lattice is $L=\Syz(U)=\{\f\in\Zx^m\,|\, \overline{U}\f =0\}$,
 where $\overline{U}=[\bu_1,\ldots,\bu_m]$ is the matrix with columns $\bu_i$.
\end{lemma}
\proof
$L$ is clearly a toric $\Zx$-lattice. Then it
suffices to show that $\I_{U} =
\I_L$, where  $\I_{U}$ is defined in \bref{eq-atv0}.
%
For $\f \in L$, we have
 $(\Y^{\f}-1)(\T^{U}) = (\T^{U})^{\f} - 1 =
 \T^{\overline{U}\f}-1 = 0$.
As a consequence, $(\Y^{\f ^{+}}-\Y^{\f^{-}})(\T^{U})=0$ and
$\Y^{\f ^{+}}-\Y^{\f^{-}} \in \I_{U}$. Since
$\I_L$ is generated by $\Y^{\f ^{+}}-\Y^{\f^{-}}$ for
$\f\in L$, we have $ \I_L\subset\I_{U}$.

To prove the other direction, consider a total order $<$ for the $\sigma$-monomials
$\{\Y^\f, \f\in \N[x]^m\}$, which extends to a total order over
$\F\{\Y\}$ by comparing the largest $\sigma$-monomial in a
$\sigma$-polynomial.
We will prove $\I_{U} \subset\I_L$. Assume the
contrary, and let $f=\Sigma_{i}a_{i}\Y^{\f_{i}}\in\I_{\U}$ be a
minimal element in $\I_{U}\setminus\I_L$ under
the above order. Let $a_0\Y^{\g}$ be the biggest $\sigma$-monomial
in $f$. From $f\in\I_{U}$, we have $f(\T^{U})=0$. Since
$\Y^{\g}(\T^{U})=\T^{\overline{U}\g}$ is a $\sigma$-monomial about $\T$ and $f(\T^{U})=0$,
there exists another $\sigma$-monomial $b_0\Y^{\h}$ in $f$ such that
$\Y^{\h}(\T^{U})=\Y^{\g}(\T^{U})$. As a consequence,
$(\Y^{\g}-\Y^{\h})(\T^{U})=\T^{\overline{U}\h}(\T^{\overline{U}(\g-\h)}-1)=0$, from
which we deduce $\g-\h\in L$ and hence $\Y^{\g}-\Y^{\h} \in
\I_{U}\cap\I_L$.
Then $f-a_{0}(\Y^{\g}-\Y^{\h})\in
\I_{U}\setminus\I_L$, which contradicts to the
minimal property of $f$, since  $f-a_{0}(\Y^{\g}-\Y^{\h})<f$.\qedd

%
Let $L\subset \Zx^m$ be a $\Zx$-lattice. Define the {\em orthogonal complement} of $L$ to be
 $$ L^C := \{\f\in\Zx^m\,|\, \forall \g\in L, \langle \f,\g\rangle=0 \}$$
where $\langle \f,\g\rangle=\f^\tau\cdot \g$ is the dot product of
$\f$ and $\g$. It is easy to show that
\begin{lemma}\label{lm-tvi2}
Let $A_{m\times r}$ be a matrix representation for $L$. Then $L^C
=\ker(A^\tau)= \{\f\in\Zx^m\,|\, A^\tau \f=0\}$ and hence $\rk(L^C) =
m - \rk(L)$.
Furthermore,
if $L$ is a toric $\Zx$-lattice, then $L= (L^C)^C$.
\end{lemma}



The following lemma shows that the inverse of Lemma \ref{lm-tvi1} is
also valid.
\begin{lemma}\label{lm-tvi4}
If $\I$ is a toric $\sigma$-ideal in $k\{\Y\}$, then $\V(\I)$ is a
toric $\sigma$-variety.
\end{lemma}
\proof  Since $\I$ is a toric $\sigma$-ideal, then the $\Z[x]$-lattice corresponding to $\I$, denoted by $L$, is toric. Suppose $V=\{\bv_{1}, \ldots, \bv_{n}\}\subset \Z[x]^m$ is a set of generators of $L^C$. Regard $V$ as a matrix with columns $\bv_i$ and let $U=\{\bu_1,\ldots,\bu_m\}\subset \Z[x]^n$ be the set of rows of $V$. Consider the toric $\sigma$-variety  $X_U$ defined by the $U$. To prove the lemma, it suffices to show $X_U=\V(\I)$ or $\I_U=\I$. Since toric $\D$-ideals and toric $\Z[x]$-lattices are in a one-to-one correspondence, we only need to show $\Syz(U)=L$. This is clear since $\Syz(U)=\ker(V)=(L^C)^C=L$.
\qedd

\begin{example}\label{ex-tvi11}
Use notations introduced in  Example \ref{ex-tv1}.
Let $\f_1=(1-x,2,0,0)^\tau,\f_2=(0,0,1-x,2)^\tau$. Then
$L=\ker(M) = (\f_1,\f_2)_{\Zx}\subseteq \Z[x]^4$. By Lemma
\ref{lm-tvi1},
we have
 $\I_U=\I_L=[y_1y_2^{2}-y_1^x, y_{3}y_{4}^{2}-y_{3}^x].$
Conversely, let $L=(\f_1,\f_2)_{\Zx}$ be the support lattice
of $\IH(X_{U})$. Then $M^{\tau}$ is the defining matrix for
$L^C$. By Lemma \ref{lm-tvi4}, $M$ is the defining matrix
for the toric $\sigma$-variety $X_{U}$.
In Example \ref{ex-tv1}, we need to use
the difference characteristic set method to compute $\I_{U}$.
Here, the only operation used to compute $\I_{U}$
is Gr\"obner basis methods for $\Zx$-lattices \cite{ghnf}.
\end{example}

Finally, we have the following effective version of Theorem \ref{th-tvi}.
\begin{theorem}\label{th-etvi}
A toric variety $X$ has the parametric representation $X_U$
and the implicit representation $\I_L$, where $U$ is given in \bref{eq-U}
and $L=(\fb)_{\Zx}$ for
$\fb=\{\f_1,\ldots,\f_s\}\subset\Zx^m$.
Then, there is a polynomial-time algorithm to compute $U$ from $F$
and vise versa.
\end{theorem}
\proof
The proofs of Lemma \ref{lm-tvi1} and  Lemma
\ref{lm-tvi2} give algorithms to compute $F$ from $U$,
and vice versa, provided we know how to compute
a set of generators of $\Syz(A)$ for a matrix $A$ with entries in $\Zx$.
In \cite{ghnf}, a polynomial-time algorithm to
compute the Gr\"obner basis for $\Zx$-lattices is given.
Combining this with  Schreyer's Theorem on page 224 of~\cite{cox-1998},
we have an algorithm to compute a Gr\"obner basis for
$\Syz(A)$ as a $\Zx$-module. Note that, when a Gr\"obner basis of the $\Zx$-lattice generated
by the columns of $A$ is given,
the complexity to compute a Gr\"obner basis of $\Syz(A)$ using Schreyer's Theorem
is clearly polynomial.\qedd

In other words, toric $\sigma$-varieties are unirational $\sigma$-varieties,
and we have efficient
implicitization and parametrization algorithms for them.

\section{$\sigma$-torus and toric $\sigma$-variety in terms of group action}
In this section, we will define the $\sigma$-torus and give another
description of toric $\sigma$-varieties in terms of group actions by
$\sigma$-tori.
%

Let $T_U^*$ be the quasi $\sigma$-torus and $X_U$ the toric $\sigma$-variety defined by $U\subset\Zx^n$ in \bref{tvsm-equ1}. In the algebraic case, $T_U^*$ is a variety, that is, $T_U^*=X_U\cap(\CN^*)^m$, where $\CN$ is the field of complex numbers and $\CN^*=\CN\setminus\{0\}$. The following example shows that this is not valid in the
$\sigma$-case.
\begin{example}\label{ex-dt1}
In Example \ref{ex-tv1},  $X_U = \V({\{y_1y_2^{2}-y_1^x,y_{3}y_{4}^{2}-y_{3}^x\}})$.
Let $P= (-1,1,-1,-1)\in \CN^4$. Then $P\in X_U(\CN)$. On the other hand, assume $P\in T_U^*(\CN)$ which means
$((t_1)^{2}$,$(t_1)^{x-1}$,\\$(t_2)^{2}$,$(t_2)^{x-1})$ $=(-1,1,-1,-1)$ or the
$\sigma$-equations $t_1^{2}+1=0,t_1^{x}-t_1=0,t_2^{2}+1=0,t_2^{x}+t_2=0$ have a solution in $(\CN^*)^2$. In what below, we will show that this is impossible.
That is, $T_U^*\varsubsetneq X_U\cap(\CN^*)^4$.

Let $\I=[t_1^2+1, t_1^x-t_1, t_2^2+1,t_2^x+t_2]$.
We have $t_2^2-t_1^2=t_2^2+1-(t_1^2+1)\in\I$.
Then, $\V(\I) = \V(\I\cup\{t_2-t_1\})\cup\V(\I\cup\{t_2+t_1\})$.
Since $t_2^x+t_2 -(t_2-t_1)^x - (t_2-t_1) - (t_1^x-t_1) = 2t_1$.
Then $\V(\I\cup\{t_2-t_1\})=\V(\I\cup\{t_2-t_1,t_1\})=\emptyset$.
Similarly, $\V(\I\cup\{t_2+t_1\})=\emptyset$ and hence $\V(\I)=\emptyset$.
\end{example}

In order to define the $\sigma$-torus, we need to introduce the
concept of Cohn $*$-closure.  $(\A^*)^m$ is isomorphic to the
$\sigma$-variety defined by
$\I_0=[y_1z_1-1,\ldots,y_mz_m-1]\subset k\{\Y,\Z\}$ in $(\A)^{2m}$,
where $\Z=(z_1,\ldots,z_m)$ is a set of $\sigma$-indeterminants.
Furthermore, $\sigma$-varieties in $(\A^*)^m$ are in a one-to-one
correspondence with affine $\sigma$-varieties contained in
$\V({\I_0})$ via the map
\begin{equation}\label{eq-TH}
\theta\colon(\A^*)^m\longrightarrow (\A)^{2m}
\end{equation}
defined by $\theta(a_1,\ldots,a_m)
=(a_1,\ldots,a_m,a_1^{-1},\ldots,a_m^{-1})$.
Let $V\subset(\A^*)^m$ and $V_1$ the Cohn closure of $\theta(V)$ in
$(\A)^{2m}$. Then $\theta^{-1}(V_1)$ is called the {\em Cohn
$*$-closure} of $V$.

Example \ref{ex-dt1} gives the motivation for the following
definition.
\begin{definition}
A {\em $\sigma$-torus} is a $\sigma$-variety which is isomorphic to the
Cohn $*$-closure of a quasi $\sigma$-torus in $(\A^{*})^{m}$.
\end{definition}

\begin{lemma}\label{lm-tv41}
Let $T_U^*$ be the quasi $\sigma$-torus defined by $U$, $T_{U}$ the Cohn $*$-closure
of $T_{U}^*$ in $(\A^*)^m$, and $\I_{U}$ defined in
\bref{eq-atv0}. Then $T_{U}$ is isomorphic to
$\Spec^{\sigma}(k\{\Y,\Z\}/\widetilde{\I}_{U})$ where
$\Z=\{z_1,\ldots,z_m\}$ is a set of $\sigma$-indeterminates and
$\widetilde{\I}_{U} = [\I_{U},y_1z_1-1,\ldots,y_mz_m-1]$
in $k\{\Y,\Z\}$.
\end{lemma}
\proof Let $\theta$ be defined in \bref{eq-TH}.
Let $\widetilde{T}_{U}^* = \theta(T_{U}^*)\subset\A^{2m}$
and $\widetilde{T}_{U}$ the Cohn closure of
$\widetilde{T}_{U}^*$ in $\A^{2m}$. Then $T_{U} =
\theta^{-1}(\widetilde{T}_{U})$ is the Cohn $*$-closure of
$T_{U}^*$ in $(\A^*)^n$. Since $\theta$ is clearly an
isomorphism between $\widetilde{T}_{U}$ and $T_{U}$,
it suffices to show that
$\IH(\widetilde{T}_{U})=\widetilde{\I}_{U}$.

We have  $\IH(\widetilde{T}_{U}) = \{f\in k\{\Y,\Z\}\mid
f(\T^{\bu_1},
\ldots,\T^{\bu_m},\T^{-\bu_1},\ldots,\T^{-\bu_m})=0\}$.
It is clear that $\widetilde{\I}_{U}\subset
\IH(\widetilde{T}_{U})$.
If $f\in \IH(\widetilde{T}_{U})$, eliminate $z_1,\ldots,z_m$ from $f$ by
replacing the
$z_i$ by $\frac{1}{y_i}$ and clear the denominates, we have $f_1= \prod_{i=1}^m y_i^{t_i} f
+f_0$, where $f_0\in \I_0$.
Substituting $y_i$ by $\T^{\bu_i}$ and $z_i$ by
$\T^{-\bu_i}$, we have $f_1(\T^{{\bu_1}}, \ldots,
\T^{{\bu_m}})=0$, and $f_1\in \I_{U}$ follows.
Then $\prod_{i=1}^m y_i^{t_i} f\in\widetilde{\I}_{U}$ and
hence $\prod_{i=1}^m z_i^{t_i}y_i^{t_i} f = \prod_{i=1}^m(y_iz_i-1+1)^{t_i} f = f
+f_0 \in\widetilde{\I}_{U}$, where $f_0 \in\I_0$. Thus
$f\in\widetilde{\I}_{U}$.
\qedd


\begin{cor}\label{cor-tv42}
Let $T_{U}$ and $X_{U}$ be the $\sigma$-torus and
the toric $\sigma$-variety defined by ${U}$, respectively.
Then $T_{U} = X_{U}\cap (\A^*)^m$. As a consequence,
$T_{U}$ is a Cohn open subset of $X_{U}$.
\end{cor}
\proof From Lemma \ref{lm-tv41} and the fact $\I_{U} =
\IH(X_{U})$, we have $T_{U} = X_{U}\cap
(\A^*)^m$.
\qedd

\begin{theorem}\label{th-TT}
Let $T$ be an affine $\sigma$-variety. Then $T$ is a $\sigma$-torus if and only if there exists a $\Z[x]$-lattice $L$ such that $T\simeq \Spec^{\sigma}(k[L])$.
\end{theorem}
\proof
We follow the notations in Lemma \ref{lm-tv41}.
Suppose $T$ is defined by $U$ and let $L=(U)_{\Z[x]}$. Since $T\simeq \widetilde{T_U}$, we just need to show
the $\D$-coordinate ring of $\widetilde{T_U}$ is $k[L]$. By definition, $\widetilde{T_U}$ is the toric $\D$-variety defined by $U\cup(-U)$. Thus the affine $\N[x]$-semimodule corresponding to $\widetilde{T_U}$ is $\N[x](U\cup(-U))=L$ and hence the $\D$-coordinate ring of $\widetilde{T_U}$ is $k[L]$.
Conversely, suppose $L=(U)_{\Z[x]}$ and $U$ is a finite subset of $\Z[x]^n$. Then by the proof of the above necessity, $U$ defines a $\sigma$-torus $T_U$ whose $\D$-coordinate ring is $k[L]$. Since $T\simeq T_U$, $T$ is a $\D$-torus.
\qedd

As a consequence, a $\sigma$-torus is also a toric $\sigma$-variety.
An algebraic torus is isomorphic to $(\CH^*)^m$ for some $m\in\N$
\cite{cox-2010}. The following example shows that this is not valid
in the  difference case.
\begin{example}\label{ex-ext1}
Let $\bu_1 = (2)$, $\bu_2 = (x)$, and $U =\{\bu_1,\bu_2\}$.
We claim that $T_U$ is not
isomorphic $\AH^*$. By Theorem \ref{th-TT}, we need to show
$\E_1=k\{t,t^{-1}\}$ is not isomorphic to
$\E_2=k\{s^2,s^{-2},s^x,s^{-x}\}$, where $t$ and $s$ are
$\sigma$-indeterminates.
Suppose the contrary, there is an isomorphism $\theta:
\E_1\Rightarrow\E_2$ and $\theta(t)=p(s)\in \E_2$.
Then there exists a $q(z)\in k\{z\}$ such that $s^2=q(p(s))$ which
is possible only if $q=z,p=s^2$.
Since $s^x\in\E_2$, there exists an $r(z)\in k\{z\}$ such that
$s^x=r(s^2)$ which is impossible.
\end{example}

Suppose $S$ is an affine $\N[x]$-semimodule. Let $(S)_{\Z[x]}$ be the $\Z[x]$-lattice generated by $S$. Let $X=\Spec^{\sigma}(k[S])$ and $T=\Spec^{\sigma}(k[(S)_{\Zx}])$.
Following Proposition \ref{prop-oo},
let $\gamma\colon S\rightarrow K$ be an element of $X(K)$ which lies in $T(K)$.
Since elements of $T(K)$ are invertible, $\gamma(S)\subseteq K^*$ and hence $\gamma$ can be extended to $\widetilde{\gamma}\colon (S)_{\Zx}\rightarrow K^*$. Similar to Proposition \ref{prop-oo},
we have

\begin{prop}
There is a one-to-one correspondence between $T(K)$ and $\Hom((S)_{\Zx},K^*)$, $\forall K\in \mathscr{E}_k$. Equivalently, $T\simeq \Hom((S)_{\Zx}, (\A^*)^1)$.
%
So we can identity an element of $T(K)$ with a morphism from $(S)_{\Zx}$ to $K^*$.
\end{prop}

A $\sigma$-variety $G$ is called a {\em $\sigma$-algebraic group} if
$G$ has a group structure and the group multiplication and the inverse map are both morphisms of $\sigma$-varieties \cite{wibmer-group}.
\begin{lemma}\label{cor-tv43}
A $\sigma$-torus $T$ is a $\sigma$-algebraic group.
\end{lemma}
\proof
For $\phi,\psi\in T$, define $\phi\cdot\psi=\phi\psi$. It is easy to check $T(K)$ becomes a group under the multiplication for each $K\in \mathscr{E}_k$. Note if $T\subseteq (\A^*)^m$, the group multiplication of $T$ is just the usual termwise multiplication of $\A^m$, namely, $\forall (x_1,\dots,x_m),(y_1,\ldots,y_m)\in T, (x_1,\dots,x_m)\cdot(y_1,\ldots,y_m)=(x_1y_1,\dots,x_my_m)$. So it is obviously a morphism of $\sigma$-varieties and so is the inverse map due to \bref{eq-TH}. Therefore, $T$ is a $\sigma$-algebraic group.
\qedd

We interpret what is a $\sigma$-algebraic group action on a $\sigma$-variety.
\begin{definition}
Let $G$ be a $\sigma$-algebraic group and $X$ a $\sigma$-variety. We say $G$ has a {\em $\sigma$-algebraic group action} on $X$ or $G$ acts on $X$ $\sigma$-algebraically if there exists a morphism of $\sigma$-varieties
\[\phi\colon G\times X\longrightarrow X\]
such that for any $ K\in \mathscr{E}_k$,
\[\phi_K\colon G(K)\times X(K)\longrightarrow X(K)\]
is a group action of $G(K)$ on $X(K)$, that is $\phi_K(1,x)=x$ and $\phi_K(g_1\cdot g_2,x)=\phi_K(g_1,\phi_K(g_2,x))$,\\$\forall x\in X(K), \forall g_1,g_2\in G(K)$.
\end{definition}

%
%

The following theorem gives a description of toric $\sigma$-varieties in terms of  group actions.
\begin{theorem}\label{aga-thm1}
A $\sigma$-variety $X$ is toric if and only if $X$ contains a
$\sigma$-torus $T$ as an open subset and with a $\sigma$-algebraic group action of
$T$ on $X$ extending the natural $\sigma$-algebraic group action of $T$ on
itself.
%
\end{theorem}
\proof
 $``\Rightarrow"$ By Corollary \ref{cor-tv42}, $T_U$ is an open subset of $X_U$.
 By Lemma \ref{cor-tv43}, $T_U$ is a $\sigma$-algebraic group.
To show that $T_U$ acts on $X_U$ as a $\sigma$-algebraic group,
define a map $X\times X\rightarrow X\colon (x_1,\ldots,x_m)\cdot (y_1,\ldots,y_m)=(x_1y_1,\ldots,x_my_m)$. It can be described using $\N[x]$-semimodule morphisms as follows: for each $K\in \mathscr{E}_k$, let $\phi,\psi \colon S\rightarrow K$ be two elements of $X(K)$, then $(\phi,\psi)\mapsto \phi \cdot \psi \colon S\rightarrow K,\phi \cdot \psi(\bu)=\phi(\bu)\cdot\psi(\bu),\forall \bu\in S$. This corresponds to the $k$-$\sigma$-algebra homomorphism $\Phi\colon k[S]\rightarrow k[S]\otimes k[S]$ such that $\Phi(\T^{\bu})=\T^{\bu}\otimes\T^{\bu},\forall \bu\in S$.
Via the embedding $T\subseteq X$, the operation on $X$ induces a map $T\times X\rightarrow X$ which is clearly a $\sigma$-algebraic group action on $X$ and extends the group action of $T$ on itself.

 $``\Leftarrow"$
There is a $\mathbb{Z}[x]$-lattice $L$ such that $T\simeq \Spec^{\sigma}(k[L])$. The open immersion $T\subseteq X$ induces $k\{X\}\subseteq k[L]$. Since the action of $T$ on itself extends to a $\sigma$-algebraic group action on $X$, we have the following commutative diagram:
\begin{equation}\label{atv-diag1}
\begin{gathered}
\xymatrix{T\times T \ar[r]^(0.6){\phi}\ar[d]&T\ar[d]\\T\times X\ar[r]^(0.6){\widetilde{\phi}}&X}
\end{gathered}
\end{equation}
where $\phi$ is the group action of $T$, $\widetilde{\phi}$ is the extension of $\phi$ to $T\times X$.
From (\ref{atv-diag1}), we obtain the following commutative diagram of corresponding $\sigma$-coordinate rings:
$$\begin{gathered}
\xymatrix{k\{X\}\ar[r]^(0.35){\widetilde{\Phi}}\ar[d]&k[L]\otimes_k k\{X\}\ar[d]\\k[L]\ar[r]^(0.35){\Phi}&k[L]\otimes_k k[L]}
\end{gathered}$$
where the vertical maps are inclusions, and $\Phi(\T^{\bu})=\T^{\bu}\otimes \T^{\bu}$ for $\bu\in L$. It follows that if $\sum_{\bu\in L}\alpha_{\bu}\T^{\bu}$ with finitely many $\alpha_{\bu}\neq 0$ is in $k\{X\}$, then $\sum_{\bu\in L}\alpha_{\bu}\T^{\bu}\otimes \T^{\bu}$ is in $k[L]\otimes_k k\{X\}$, so $\alpha_{\bu}\T^{\bu}\in k\{X\}$ for every $\bu\in L$. This shows that there is a subset $S$ of $L$ such that $k\{X\}=k[S]=\bigoplus_{\bu\in S}k\T^{\bu}$. Since $k\{X\}$ is a $k$-$\sigma$-subalgebra of $k[L]$, it follows that $S$ is an $\mathbb{N}[x]$-semimodule. And since $k\{X\}$ is a finitely $\sigma$-generated $k$-$\sigma$-algebra, $S$ is finitely generated, thus it is an affine $\mathbb{N}[x]$-semimodule. So by Theorem \ref{astv-thm1}, $X$ is a toric $\D$-variety.

\qedd

\section{Toric $\sigma$-varieties and affine $\N[x]$-semimodules}
In this section, deeper connections between
toric $\sigma$-varieties and affine $\N[x]$-semimodules will be established.
%
We first show that
the category of toric $\sigma$-varieties with toric morphisms is antiequivalent to the category of affine $\N[x]$-semimodules with $\N[x]$-semimodule morphisms.

If $\phi\colon S\rightarrow S'$ is a morphism between two affine $\mathbb{N}[x]$-semimodules, we have an induced $k$-$\sigma$-algebra homomorphism $f\colon k[S]\rightarrow k[S']$ such that $f(\T^{\bu})=\T^{\phi({\bu})},\bu\in S$.
\begin{definition}
Let $X_i=\Spec^{\sigma}(k[S_i])$ be the toric $\sigma$-varieties coming from affine $\N[x]$-semimodules $S_i, i=1,2$ with $\D$-torus $T_i$ respectively. A morphism $\phi\colon X_1\rightarrow X_2$ is called {\em toric} if $\phi(T_1)\subseteq T_2$ and $\phi|_{T_1}$ is a $\sigma$-algebraic group homomorphism.
\end{definition}

\begin{prop}
Let $\phi\colon X_1\rightarrow X_2$ be a toric morphism of toric $\sigma$-varieties. Then $\phi$ preserves group actions, namely,
$\phi(t\cdot p)=\phi(t)\cdot \phi(p)$
for all $t\in T_1$ and $p\in X_1$.
\end{prop}
\proof
Suppose the action of $T_i$ on $X_i$ is given by a morphism $\varphi_i\colon T_i\times X_i\rightarrow X_i, i=1,2$. Preserving group action means the following diagram is commutative:
\begin{equation}
\begin{gathered}
\xymatrix{T_1\times X_1\ar[r]^(0.6){\varphi_1}\ar[d]_{\phi|_{T_1}\times\phi}&X_1\ar[d]^{\phi}\\T_2\times X_2\ar[r]^(0.6){\varphi_2}&X_2}
\end{gathered}
\end{equation}
If we replace $X_i$ by $T_i$ in the diagram, then it certainly commutes since $\phi|_{T_1}$ is a group homomorphism. Since $T_1\times T_1$ is dense in $T_1\times X_1$, the whole diagram is commutative.
\qedd

\begin{lemma}\label{tmtv-prop}
Let $T_i=\Spec^{\sigma}(k[L_i])$ be two $\sigma$-tori defined by the $\Zx$-lattices $L_i,i=1,2$. Then a map $\phi\colon T_1\rightarrow T_2$ is a $\sigma$-algebraic group homomorphism if and only if the corresponding map of $\sigma$-coordinate rings $\phi^*\colon k[L_2]\rightarrow k[L_1]$ is induced by a $\Z[x]$-module homomorphism $\hat{\phi}\colon L_2\rightarrow L_1$.
\end{lemma}
\proof
``$\Leftarrow$''. Suppose $\hat{\phi}\colon L_2\rightarrow L_1$ is a $\Z[x]$-module homomorphism and it induces a morphism of $\sigma$-varieties $\phi\colon T_1\rightarrow T_2$ via $\phi^*$. Then, for any $\varphi,\psi\in T_1=\Hom(L_1,(\A^*)^1)$, $\phi(\varphi\cdot\psi)=(\varphi\cdot\psi)\circ\hat{\phi}=(\varphi\circ\hat{\phi})\cdot(\psi\circ\hat{\phi})
=\phi(\varphi)\cdot\phi(\psi)$. So $\phi$ is also a morphism of groups and hence a morphism of $\sigma$-algebraic groups.

``$\Rightarrow$''. Suppose $\phi\colon T_1\rightarrow T_2$ is a morphism of $\sigma$-algebraic groups. Then we have the following commutative diagram:
$$\begin{gathered}
\xymatrix{k[L_2]\ar[r]\ar[d]^{\phi^*}&k[L_2]\otimes_k k[L_2]\ar[d]^{\phi^*\otimes \phi^*}\\k[L_1]\ar[r]&k[L_1]\otimes_k k[L_1]}
\end{gathered}$$
Given $\bv\in L_2$, there is a finite subset $S$ of $L_1$ such that
$\phi^*(\T^{\bv})=\sum_{\bu\in S}\alpha_{\bu}\T^{\bu}$.
It follows from the commutativity of the diagram that $\sum_{\bu\in S}\alpha_{\bu}\T^{\bu}\otimes\T^{\bu}=\sum_{\bu_1\in S,\bu_2\in S}\alpha_{\bu_1}\alpha_{\bu_2}\T^{\bu_1}\otimes\T^{\bu_2}$. This shows that there is at most one $\bu$ with $\alpha_{\bu}\neq0$ and in this case $\alpha_{\bu}=1$. Note that $\T^{\bv}$ is invertible in the group $T_1$, so $\phi^*(\T^{\bv})\neq0$. So we have $\phi^*(\T^{\bv})=\T^{\bu}$ for some $\bu\in L_1$. Then we can define a map $\hat{\phi}\colon L_2\rightarrow L_1$, $\bv\mapsto \bu$. It is easy to check that $\hat{\phi}$ is a $\Z[x]$-module homomorphism.
\qedd

\begin{lemma}\label{tmtv-thm}
Let $X_i=\Spec^{\sigma}(k[S_i])$ be toric $\sigma$-varieties coming from affine $\N[x]$-semimodules $S_i, i=1,2$ with $\D$-torus $T_i$ respectively. Then a morphism $\phi\colon X_1\rightarrow X_2$ is toric if and only if it is induced by an $\N[x]$-semimodule morphism $\hat{\phi}\colon S_2\rightarrow S_1$.
\end{lemma}
\proof
``$\Leftarrow$''. Suppose $\hat{\phi}\colon S_2\rightarrow S_1$ is an $\N[x]$-semimodule morphism.
Then $\hat{\phi}$ extends to a $\Z[x]$-module homomorphism $\hat{\phi}\colon L_2\rightarrow L_1$, where $L_1=(S_1)_{\Zx}, L_2=(S_2)_{\Zx}$.
By Lemma \ref{tmtv-prop}, it induces a morphism of $\sigma$-algebraic groups $\phi\colon T_1\rightarrow T_2$. So $\phi$ is toric.

``$\Rightarrow$''. Since $\phi$ is toric, $\phi|_{T_1}$ is a $\sigma$-algebraic group homomorphism. By Lemma \ref{tmtv-prop}, it is induced by a $\Z[x]$-module homomorphism $\hat{\phi}\colon L_2\rightarrow L_1$. This, combined with $\phi^*(k[S_2])\subseteq k[S_1]$, implies that $\hat{\phi}$ induces an $\N[x]$-semimodule morphism $\hat{\phi}\colon S_2\rightarrow S_1$.
\qedd

%
Combining Theorem \ref{astv-thm1} and Lemma \ref{tmtv-thm}, we have
\begin{theorem}
The category of toric $\sigma$-varieties with toric morphisms is antiequivalent to the category of affine $\N[x]$-semimodules with $\N[x]$-semimodule morphisms.
\end{theorem}

In the rest of this  section, we establish a one-to-one correspondence between irreducible $T$-invariant subvarieties of a toric $\sigma$-variety and faces of the corresponding affine $\N[x]$-semimodule. A one-to-one correspondence between $T$-orbits of a toric $\sigma$-variety and faces of the corresponding affine $\N[x]$-semimodule for a class of $\N[x]$-semimodules.

\begin{definition}
Let $S$ be an affine $\mathbb{N}[x]$-semimodule. Define a {\em face} of $S$ to be an $\mathbb{N}[x]$-subsemimodule $F\subseteq S$ such that
\begin{description}
\item[(1)] $\forall \BU_1,\BU_2\in S$, $\BU_1+\BU_2\in F$ implies $\BU_1,\BU_2\in F$;
\item[(2)]  $\forall \BU\in S$, $x\bu\in F$ implies $\BU\in F$,
\end{description}
which is denoted by  $F\preceq S$.
\end{definition}

Note if $S=\mathbb{N}[x](\{\BU_1,\BU_2,\ldots,\BU_m\})$ and $F$ is a face of $S$, then $F$ is generated by a subset of $\{\BU_1,\BU_2,\ldots,\BU_m\}$ as an $\mathbb{N}[x]$-semimodule. It follows that $F$ is an affine $\mathbb{N}[x]$-semimodule and $S$ has only finitely many faces. $S$ is a face of itself.
It is easy to prove that the intersection of two faces is again a face and a face of a face is again a face. $S$ is called {\em pointed} if $S\cap (-S)=\{0\}$, i.e.\ $\{0\}$ is a face of $S$.


\begin{example}
Let $S=\N[x](\{\BU_1=(x,1),\BU_2=(x,2),\BU_3=(x,3)\})$. Then $S$ has four faces: $F_1=\{0\}$, $F_2=\N[x](\BU_1)$, $F_3=\N[x](\BU_3)$ and $F_4=S$. Since $2\BU_2=\BU_1+\BU_3$, $\BU_2$ does not generate a face.
\end{example}

For an $\N[x]$-semimodule $S\subset\Z[x]^n$, a {\em $\sigma$-monomial} in $k[S]$ is an element of the form $\T^{\bu}$ with $\bu\in S$. If we define a degree map by $\deg(\T^{\bu})=\bu, \forall \bu\in S$, then $k[S]$ becomes a $S$-graded ring. A $\sigma$-ideal of $k[S]$ is called {\em $S$-homogeneous} if it can be generated by homogeneous elements, i.e.\ $\sigma$-monomials.

\begin{lemma}\label{oatv-lemma2}
A subset of $F$ of $S$ is a face if and only if $k[S\backslash F]$ is a $\sigma$-prime ideal of $k[S]$.
\end{lemma}
\proof
Let $I=k[S\backslash F]$. Since $I$ is $S$-homogeneous, we just need to consider homogeneous elements, that is $\sigma$-monomials (\cite[Propsition 3.6]{Jie}). The condition for $I$ to be a $\sigma$-ideal is equivalent with the fact that whenever $\BU_1\in S\backslash F$ or $\BU_2\in S\backslash F$, then $\BU_1+\BU_2\in S\backslash F$ and $\forall \BU\in S\backslash F$, then $x\bu\in S\backslash F$, i.e.\ $\BU_1+\BU_2\in F\Rightarrow \BU_1,\BU_2\in F$ and $x\bu\in F\Rightarrow \BU\in F$. Moreover, the condition for $I$ to be $\sigma$-prime is equivalent with the fact that if $\BU_1+\BU_2\in S\backslash F$, then $\BU_1\in S\backslash F$ or $\BU_2\in S\backslash F$ and if $x\bu\in S\backslash F$, then $\BU\in S\backslash F$, i.e.\ $\BU_1,\BU_2\in F\Rightarrow \BU_1+\BU_2\in F$ and $\BU\in F\Rightarrow x\bu\in F$. So $F$ is a face if and only if $I$ is a $\sigma$-prime ideal.
\qedd

Let $X=\Spec^{\sigma}(k[S])$ be a toric $\sigma$-variety and $T$ the $\sigma$-torus of $X$.
A $\sigma$-subvariety $Y$ of $X$ is called {\em invariant} under the action of $T$
if $T\cdot Y\subset Y$. The following theorem gives a description for invariant
irreducible invariant $\sigma$-subvarieties of $X$.
\begin{theorem}
The irreducible invariant $\sigma$-subvarieties of $X$ under the action of $T$ are in an inclusion-preserving bijection with the faces of $S$. More precisely, if we denote the irreducible invariant $\sigma$-subvariety corresponding to the face $F$ by $D(F)$, then $D(F)$ is defined by the $\sigma$-ideal $k[S\backslash F]=\bigoplus_{\BU\in S\backslash F}k\T^{\BU}$ and the $\sigma$-coordinate ring of $D(F)$ is $k[F]=\bigoplus_{\BU\in F}k\T^{\BU}$.
\end{theorem}
\proof
Let $L=(S)_{\Zx}$. Suppose $Y$ is an irreducible $\sigma$-subvariety of $X$ and is defined by the $\sigma$-ideal $I$. Then $k\{Y\}=k[S]/I$. By definition, $Y$ is invariant under the $\sigma$-torus action if and only if the action of $T$ on $X$ induces an action on $Y$, that is, we have the following commutative diagram:
$$\begin{gathered}
\xymatrix{k[S]\ar[r]^(0.37){\phi}\ar[d]&k[L]\otimes k[S]\ar[d]\\k\{Y\}\ar[r]&k[L]\otimes k\{Y\}}
\end{gathered}$$
Since $k[L]\otimes k\{Y\}=k[L]\otimes (k[S]/I)\simeq k[L]\otimes k[S]/k[L]\otimes I$, this is the case if and only if $\phi(I)\subseteq k[L]\otimes I$. As in the proof of Theorem \ref{aga-thm1}, this is equivalent with the fact that $I$ is an $L$-graded ideal of $k[S]$, that is,  we can write $I=\oplus_{\BU\in S'}k\T^\BU$, where $S'$ is a subset of $S$.
%
%
By Lemma \ref{oatv-lemma2}, $I$ is a $\sigma$-prime ideal $\iff F=S\backslash S'$ is a face of $S$. Moreover, since $I=k[S\backslash F]$, $k\{Y\}=k[S]/I=k[F]$.
\qedd

Note that an element $\gamma\colon S\rightarrow K$ of $X(K)$ which lies in $D(F)(K)$ if and only if $\gamma(S\backslash F)=0$.

Suppose $X$ is a toric $\sigma$-variety with $\D$-torus $T$. By Theorem \ref{aga-thm1}, for each $K\in \mathscr{E}_k$, $T(K)$ has a group action on $X(K)$, so we have orbits of $T(K)$ in $X(K)$ under the action. To construct a correspondence between orbits and faces, we need a new kind of affine $\N[x]$-semimodules.
An affine $\N[x]$-semimodule $S$ is said to be {\em face-saturated} if for any face $F$ of $S$, a morphism $\phi\colon F\rightarrow K^*$ can be extended to a morphism $\widetilde{\phi}\colon S\rightarrow K^*$. A necessary condition for $S$ to be face-saturated is that for any face $F$ of $S$, $(F)_{\Z[x]}$ is $\N[x]$-saturated in $(S)_{\Z[x]}$, that is, for any $g\in\N[x]\backslash\{0\}$, $\BU\in (S)_{\Z[x]}$, $g\bu\in (F)_{\Z[x]}$ implies $\BU\in (F)_{\Z[x]}$.

\begin{example}
Let $S=\N[x](\{(2,0),(1,1),(0,1)\})$ and $F=\N[x](\{(2,0)\})$ a face of $S$. $(1,0)\in (S)_{\Zx}$. Since $(1,0)\notin F$ and $2(1,0)\in F$, $S$ is not face-saturated.
\end{example}

Now we prove the following Orbit-Face correspondence theorem.
\begin{theorem}\label{oatv-thm}
Suppose $S$ is a face-saturated affine $\N[x]$-semimodule. Let $X=\Spec^{\sigma}(k[S])$ be the toric $\sigma$-variety of $S$ and $T$ the $\sigma$-torus of $X$. Then for each $K\in \mathscr{E}_k$, there is a one-to-one correspondence between the faces of $S$ and the orbits of $T(K)$ in $X(K)$.
\end{theorem}
\proof
Suppose $F$ is a face of $S$. The inclusion $F\subseteq S$ induces a morphism of toric $\sigma$-varieties $f\colon X\rightarrow Y$ and a morphism of $\sigma$-tori $g\colon T\rightarrow T_Y$, where $Y=\Spec^{\sigma}(k[F])$ and $T_Y=\Spec^{\sigma}(k[(F)_{\Zx}])$. For each $K\in \mathscr{E}_k$, the elements of $T_Y(K)$ are such morphisms $\phi\colon S\rightarrow K$ such that $\phi(F)\subseteq K^*$ and $\phi(S\backslash F)=0$. Since $S$ is face-saturated, they can be extended to morphisms $\widetilde{\phi}\colon S\rightarrow K^*$ which are elements of $T(K)$. So $g_K\colon T(K)\rightarrow T_Y(K)$ is surjective. Suppose $\psi\colon S\rightarrow K^*$ is an element of $T(K)$, then the action of $\psi$ on $\phi$ is $\psi\phi$ which is still an element of $T_Y(K)$. So $T_Y(K)$ is closed under the action of $T(K)$.
Suppose $e$ is the identity element of $T_Y(K)$, then $T(K)\cdot e=g_K(T(K))\cdot e$. Since $g_K$ is surjective, $g_K(T(K))=T_Y(K)$ and $T(K)\cdot e=T_Y(K)$. Therefore, $T_Y(K)$ is transitive under the action of $T(K)$. Thus $T_Y(K)$ is an orbit for the action of $T(K)$ on $X(K)$.

On the other hand, for each $K\in \mathscr{E}_k$, given an element $\phi\colon S\rightarrow K$ in $X(K)$, let $F:=\phi^{-1}(K^*)$. Then for any $\BU_1,\BU_2\in F$ and $ g_1,g_2\in \mathbb{N}[x]$, $\phi(\BU_1g_1+\BU_2g_2)=\phi(\BU_1)^{g_1}\phi(\BU_2)^{g_2}\in K^*\Rightarrow \BU_1g_1+\BU_2g_2\in F$.
Therefore $F$ is an $\N[x]$-subsemimodule of $S$.
Moreover, for $\BU_1,\BU_2,\BU\in F$, whenever $\BU_1+\BU_2\in F\Rightarrow \phi(\BU_1+\BU_2)=\phi(\BU_1)\phi(\BU_2)\in K^*\Rightarrow \phi(\BU_1),\phi(\BU_2)\in K^*\Rightarrow \BU_1,\BU_2\in F$; whenever $x\BU\in F\Rightarrow \phi(x\bu)=\phi(\BU)^x\in K^*\Rightarrow \phi(\BU)\in K^* \Rightarrow \BU\in F$. So $F$ is a face of $S$. Let $Y=\Spec^{\sigma}(k[F]),T_Y=\Spec^{\sigma}(k[(F)_{\Zx}])$. It is clear that $\phi\in T_Y(K)$ and $T_Y(K)$ is the orbit of $\phi$ in $X(K)$.
It is clear that two different faces gives two discrete orbits, which proves the one-to-one correspondence.
\qedd

\section{An order bound of toric $\sigma$-variety}
\label{sec-ord}
In this section, we show that the $\sigma$-Chow form
\cite{dd-chowform,gao} of a toric $\sigma$-variety $X_{U}$
is the sparse $\sigma$-resultant \cite{dd-sres} with support
${U}$.
As a consequence, we can give a bound for the order of
$X_{U}$.

Let ${U}=\{\bu_1,\ldots, \bu_m\}$ be a subset of
$\Z[x]^n$ and $X_{U}$ the toric $\sigma$-variety defined by
${U}$.
In order to establish a connection between the $\sigma$-Chow form of
$X_{U}$ and the $\sigma$-sparse resultant with support
${U}$, we assume that ${U}$ is {\em Laurent
transformally essential} \cite{dd-sres}, that is $\rk(\overline{U})=n$
by regrading $\overline{U}$ as a matrix with $\bu_i$ as the $i$-th column.

Let $\T =\{t_1, \ldots, t_n\}$ be a set of $\sigma$-indeterminates.
Here, the fact that ${U}$ is Laurent transformally essential means that
there exist indices $k_1,\ldots,k_n\in\{1,\ldots,m\}$ such that
the Laurent $\sigma$-monomials
$\T^{\bu_{k_1}},\ldots,\T^{\bu_{k_n}}$ are transformally
independent over $k$ \cite{dd-sres}.

Let
$\mathcal{A}=\{M_1=\T^{\bu_{1}},\ldots,M_m=\Y^{\bu_{m}}\}$
and
 \begin{equation}\label{eq-P}
 \P_i=a_{i0}+a_{i1}M_1+\cdots+a_{im}M_m\,(i=0,\ldots,n)
 \end{equation}
$n+1$ generic Laurent $\sigma$-polynomials with the same support
$U$. Denote ${\CC_i}=(a_{i0},\ldots,a_{im}),$ $i=0,\ldots,n$.
Since $\mathcal{A}$ is Laurent transformally essential, the $\sigma$-sparse
resultant of $\P_0, \P_1,\ldots,$ $\P_n$ exists \cite{dd-sres},
which is denoted by $R_{U}\in k\{\CC_0,\ldots,\CC_n\}$.

By Lemma \ref{lm-tv10},   $X_{U}\subset \AH^m$ is an
irreducible $\sigma$-variety of dimension $\rk(U)=n$. Then, the
$\sigma$-Chow form of $X_{U}$, denoted by
$C_{U}\in k\{\CC_0,\ldots,\CC_n\}$, can be obtained by
intersecting $X_{U}$ with the following generic
$\sigma$-hyperplanes \cite{dd-chowform}
 $$\L_i=a_{i0}+a_{i1}y_1+\cdots+a_{im}y_m\,(i=0,\ldots,n).$$
We have
\begin{theorem}\label{th-toric-chow}
Up to a sign, the sparse $\sigma$-resultant $R_{U}$ of
$\P_i\,(i=0,\ldots,n)$ is the same as the $\sigma$-Chow form
$C_{U}$ of $X_{U}$.
\end{theorem}
\proof All $\sigma$-ideals in this proof are supposed to be in
$R=k\{\CC_0,\ldots,\CC_n,\Y,\T^{\pm}\}$, unless
specifically mentioned otherwise.
From \cite{dd-sres},
 $$[\P_0,\P_1,\ldots,\P_n]\cap k\{{\CC_0},\ldots,{\CC_n}\}=\sat(R_{U},R_1,\ldots,R_l)$$
is a $\D$-prime $\sigma$-ideal of codimension one in
$k\{\CC_0,\ldots,\CC_n\}$. Let $\I_{U} = \IH(X_U)$. From \cite{dd-chowform},
 $$[\I_{U},\L_0,\L_1,\ldots,\L_n]\cap
k\{{\CC_0},\ldots,{\CC_n}\}=\sat(C_{U},C_1,\ldots,C_t)$$ is a
$\D$-prime $\sigma$-ideal of codimension one in
$k\{\CC_0,\ldots,\CC_n\}$.
By Theorem 7 of \cite{dd-sres}, in order to prove $C_{U}=R_{U}$,
it suffices to show
 $$[\P_0,\P_1,\ldots,\P_n]\cap
 k\{{\CC_0},\ldots,{\CC_n}\}=[\I_{U},\L_0,,\L_1,\ldots,\L_n]\cap
 k\{{\CC_0},\ldots,{\CC_n}\}.$$
Let $\I_{\T}=[y_1-M_1,\ldots,y_m-M_m]$. By \bref{eq-tv01},
$\I_{U}=\I_{\T}\cap k\{\Y\}$.
Then, $[\I_{U},\L_0,\ldots,\L_n]\cap
k\{{\CC_0},\ldots,{\CC_n}\}=[y_1-M_1,\ldots,y_m-M_m,\L_0,\ldots,\L_n]\cap
k\{{\CC_0},\ldots,{\CC_n}\}=[y_1-M_1,\ldots,y_m-M_m,\P_0,\ldots,\P_n]\cap
k\{{\CC_0},\ldots,{\CC_n}\}$.
Since $\P_i\in k\{\CC_0,\ldots,\CC_m,\T^{\pm}\}$ does not contain
any $y_i^{x^j}$, we have
$[y_1-M_1,\ldots,y_m-M_m,\P_0,\ldots,\P_n]\cap
k\{{\CC_0},\ldots,{\CC_n}\} = [\P_0,\ldots,\P_n]\cap
k\{{\CC_0},\ldots,{\CC_n}\}$, and the theorem is proved.\qedd

To give a bound for the order of $X_{U}$, we need to
introduce the concept of Jacobi number.
Let $M=({m}_{ij})$ be an $n\times n$ matrix with elements either in
$\N$ or $-\infty$. A {\em diagonal sum} of $M$ is any sum
${m}_{1\sigma(1)}+{m}_{2\sigma(2)}+\cdots+{m}_{n\sigma(n)}$ with
$\sigma$ a permutation of $1,\ldots,n$. The {\em Jacobi number} of
$M$ is the maximal diagonal sum of $M$, denoted by $\Jac(M)$
\cite{dd-sres}.

Let $U=\{{\bu}_1,\ldots, {\bu}_m\}\subset\Zx^{n}$ and
$\overline{U}=(a_{ij})_{m\times n}$ the matrix with $\bu_i$ as the $i$-th
column.
For each $i\in\{1,\ldots,n\}$, let $o_{i} = \max_{k=1}^n
\deg(a_{ik},x)$ and assume that $\deg(0,x)=-\infty$. Since $\overline{U}$ does
not contain zero rows, no $a_{ij}$ is $-\infty$.
For a $p(x)\in\Zx$, let $\underline{\deg}(p,x) = \min\{k\in\N\,|\,
s.t.\, \coeff(p,x^k)\ne0 \}$ and $\underline{\deg}(0,x)=0$.
For each $i\in\{1,\ldots,n\}$, let $\underline{o}_i=\min_{k=1}^{n}
\underline{\deg}(a_{ik},x)$ and $\underline{o}=\sum_{i=1}^n
\underline{o}_i$. Then we have
\begin{theorem}\label{th-tv-ord}
Use the notations just introduced. Let $X_{U}$ be the toric
$\sigma$-variety defined by $U$. Then $\ord(X_{U}) \le
\sum_{i=1}^n (o_i-\underline{o}_i)$.
\end{theorem}
\proof Use the notations in Theorem \ref{th-toric-chow}. Since
$\P_i$ in \bref{eq-P} have the same support for all $i$,
$\ord(R_{U},\CC_i)$ are the same for all $i$.
The {\em order matrix} for $\P_i$ given in \bref{eq-P} is
$O=(\ord(\P_i,t_j))_{(n+1)\times n}=(o_{ij})_{(n+1)\times n}$, where
$o_{ij} = o_j$. That is, all rows of $O$ are the same.
Let $\overline{O}$ be obtained from $O$ by deleting the any row of
$O$. Then $J=\Jac(\overline{O})=\sum_{i=1}^n o_i$.
By Theorem 4.17 of \cite{dd-sres}, $\ord(R_{U},\CC_i) \le J-
\underline{o} = \sum_{i=1}^n (o_i-\underline{o}_i)$.
By Theorem 6.12 of \cite{dd-chowform}, $\ord(X_{U}) =
\ord(C_{U},\CC_i)$ for each $i=0,\ldots,n$.
By Theorem \ref{th-toric-chow}, $C_{U} = R_{U}$. Then
the theorem is proved.\qedd

\section{Algorithms}\label{sec-alg}
In this section, we give algorithms to decide whether a given $\Zx$-lattice $L$
is toric and in the negative case to compute the $\Zx$-saturation of $L$.
Using these algorithms,
for a $\Zx$-lattice $L$ generated by $U = \{\bu_1,\ldots,\bu_m\} \subset \Zxn$,
we can decide whether the binomial $\sigma$-ideal
$\I_L$ is toric and in the negative case, to compute a toric $\Zx$-lattice $L' \supset L$ such that
$\I_{L'} \supset \I_L$ is the smallest toric ideal containing $\I_L$.

We first introduce the concept of  Gr\"{o}bner bases for $\Zx$-lattices.
For details, please refer to \cite{cox-1998,gao-dbi}.
Denote ${\beps}_i$ to be the $i$-th standard basis vector
$(0,\ldots,0,1,0,\ldots,0)^\tau\in\Z[x]^n$, where $1$ lies in the
$i$-th row of ${\beps}_i$. A {\em monomial} ${\bf m}$ in $\Z[x]^n$
is an element of the form $ax^k{\beps}_i\in\Z[x]^n$, where $a\in\Z$
and $k\in\N$.
The following {\em monomial order} $>$ of $\Z[x]^n$ will be used in
this paper: $ax^\alpha{\beps}_i > bx^\beta{\beps}_j$ if $i
> j$, or $i=j$ and $\alpha > \beta$, or $i=j$, $\alpha = \beta$, and
$|a|>|b|$.

For any $\f\in\Z[x]^n$, the largest monomial in $\f$ is called the {\em
leading term} of $\f$  is denoted to be $\LT(\f).$
The order $>$ can be extended to elements of $\Zxn$ as follows: for
$\f,\g\in\Zxn$, $\f <\g$ if and only if $\LT(\f) < \LT(\g)$.

A monomial $ax^\alpha{\beps}_i$ is said to be reduced w.r.t another nonzero monomial
$bx^\beta{\beps}_j$  if $i \ne j$; $i=j$, $\alpha < \beta$; $i=j$, $\alpha \ge \beta$, and $0\le a < |b|$.
Let $\G\subset\Z[x]^n$ and  $\f\in\Z[x]^n$. We say that $\f$ is {\em
reduced} with respect to $\G$ if any monomial of $\f$ is not a
multiple of $\LT(\g)$ by an element in $\Zx$ for any $\g\in\G$.

A finite set $\fb = \{\f_1,\ldots,\f_s \}\subset\Z[x]^n$ is called a
{\em Gr\"{o}bner basis} for the $\Zx$-lattice $L$ generated by $\fb$
if for any $\g\in L$, there exists an $i$, such that $\LT(\g) | \LT(\f_i)$.
A Gr\"{o}bner basis $\fb$ is called {\em reduced} if for any
$\f\in\fb$, $\f$ is reduced with respect to $\fb\setminus\{\f\}$.

Let $\fb$ be a Gr\"{o}bner basis. Then any $\f\in\Zxn$ can be
reduced to a unique normal form by $\fb$, denoted by
$\grem(\f,\fb)$, which is reduced with respect to $\fb$.

Let ${\bf f,g}\in \Zx^n,~{\bf LT(f)}=ax^k{\bf e}_i,~{\bf LT(g)}=bx^s{\bf e}_j$, $s\leq k$.
The {\em S-polynomial} of ${\bf f}$ and ${\bf g}$ is defined as follows: if $i \neq j$ then $S({\bf f,g})=0$; otherwise $S({\bf f,g})=$
\begin{equation}
\left\{
  \begin{array}{ll}
    {\bf f}-\frac{a}{b}x^{k-s}{\bf g}, & \hbox{ if } b\,|\,a; \\
    \frac{b}{a}{\bf f}-x^{k-s}{\bf g}, & \hbox{ if } a\,|\,b; \\
    u{\bf f}+vx^{k-s}{\bf g}, & \hbox{if $a\nmid b\hbox{ and }~b\nmid a,~\emph{where}~ \gcd(a,b)=ua+vb$.}
  \end{array}
\right.
\end{equation}

Then, it is known that $\fb\subset\Z[x]^n$  is a Gr\"{o}bner basis
if and only if  $\grem(S(\f_i,\f_j), \fb)=0$  for all $i,j$ \cite{cox-1998,ghnf}.
%

Next, we will give the structure for the matrix representation
for the Gr\"obner basis of a $\Zx$-lattice.
Let
{\small
\begin{equation}\label{ghf}\C=\left[\begin{array}{lllllllllll}
c_{1,1}   & \ldots & c_{1,l_1}     &c_{1,l_1+1}   &\ldots       &\ldots       &\ldots &\ldots &\ldots            &\ldots  &\ldots   \\
\ldots    & \ldots & \ldots        & \ldots       &\ldots       &\ldots       &\ldots &\ldots &\ldots            &\ldots  &\ldots \\
c_{r_1, 1}& \ldots & c_{r_1, l_1}  &c_{r_1,l_1+1} & \ldots      &\ldots       &\ldots &\ldots & \ldots           &\ldots  &\ldots \\
0         & \ldots & 0             &c_{r_1+1,1}   &    \ldots   &c_{r_1+1,l_2}&\ldots &\ldots &  \ldots          &\ldots  &\ldots  \\
\ldots    & \ldots & \ldots        & \ldots       &\ldots       &\ldots       &\ldots &\ldots &\ldots            &\ldots  &\ldots  \\
0         & \ldots & 0             & c_{r_2,1}    &    \ldots   & c_{r_2,l_2} &\ldots &\ldots & \ldots           &\ldots  &\ldots   \\
\ldots    & \ldots & \ldots        &\ldots        &    \ldots   &\ldots       &\ldots &\ldots & \ldots           &\ldots  &\ldots   \\
0         & \ldots & 0             &0             &    \ldots   &0            &\ldots &0      &c_{r_{t-1}+1,1}   &\ldots  & c_{r_{t-1}+1,l_t} \\
\ldots    & \ldots & \ldots        &\ldots        &    \ldots   &\ldots       &\ldots &\ldots &\ldots            &\ldots  & \ldots \\
0         & \ldots & 0             &0             &    \ldots   &0            &\ldots &0      &c_{r_t,1}         &\ldots  & c_{r_t,l_t} \\
\end{array}\right]_{n\times s}\end{equation}}
whose elements are in $\Zx$.
We denote by $\c_i$ to be the $i$-th column of $\C$
and $\c_{i,j}$ to be the column whose $r_i$-th element is $c_{r_i,j}$
for $i=1,\ldots,t; j=1,\ldots,l_t$.
%
Let $c_{r_i,j}$ be
 \begin{equation}\label{eq-cij}
 c_{r_i,j}=c_{i,j,0}x^{d_{ij}}+\cdots+c_{i,j,d_{ij}}.\end{equation}

\begin{defn}\label{def-ghf}
The matrix $\C$ in~\bref{ghf} is called a {\em generalized Hermite
normal form} if it satisfies the following conditions:
\begin{description}
\item[1)] $0\le d_{r_i,1} < d_{r_i,2} < \cdots < d_{r_i,l_i}$ for any $i$.

\item[2)] $ c_{r_i,l_i,0} |\cdots | c_{r_i,2,0}  | c_{r_i,1,0}$.

\item[3)]
$S(\c_{r_i,j_1},\c_{r_i,j_2})=
x^{d_{r_i,j_2}-d_{r_i,j_1}}\c_{r_i,j_1}-\frac{c_{r_i,j_1,0}}{c_{r_i,j_2,0}}\c_{r_i,j_2}$
can be reduced to zero by the column vectors of the matrix for any
$1\le i\le t, 1\le j_1<j_2\le l_i$.

\item[4)]
$\c_{i}$ is reduced w.r.t. the column vectors of the matrix
other than $\c_{i}$, for any $1\le i\le s$.
\end{description}
\end{defn}
It is proved in \cite{gao-dbi} that
$\fb = \{\f_1,\ldots,\f_s \}\subset\Z[x]^n$ is a reduced Gr\"{o}bner
basis such that $\f_1< \f_2< \cdots < \f_s$ if and only if
the matrix $[\f_1,\ldots,\f_s]$
is a generalized Hermite normal form.
From \cite{ghnf}, generalized Hermite normal form can be computed
in polynomial-time.
%

For $S\subset\Zx^n$, we use $(S)_D$ to denote the $D$-module
generated by $S$ in $D^n$, where $D=\Zx$ or $D=\Q[x]$.
When $D=\Zx$, $(S)_D$ is the $\Zx$-lattice generated by $S$.
Similarly, let $A$ be any matrix with entries in $\Zx$. We use $(A)_D$ to denote
the $D$-module generated by the column vectors of $A$.

A $\Zx$-lattice $L\subset\Zxn$ is called {\em $\Z$-saturated} if,
for any $a\in\Z^*$ and $\f\in\Z[x]^n$,  $a\f\in L$ implies $\f\in L$.
The {\em $\Z$-saturation} of $L$ is defined to be
$$\sat_{\Z}(L)=\{\f\in
\Z[x]^{n}\, |\, \exists a \in \Z^* \st a\f\in L\}.$$
We need the following algorithm from \cite{gao-dbi}.

$\bullet$ {\bf{ZFactor}}$(\C)$: for a generalized Hermite normal form $\C$,
the algorithm returns $\emptyset$ if $L=(\C)_{\Zx}$ is $\Z$-saturated,
or a finite set $S\subset\sat_{\Z}(L)\setminus(L)$.

The following algorithm checks whether
$L=(\C)_{\Zx}$ is $\Zx$-saturated and in the negative case
returns elements of $\sat_{\Zx}(L)\setminus L$.
\begin{algorithm}[H]\label{alg-D-satZX}
  \caption{\bf --- ZXFactor$(\C)$} \smallskip
  \Inp{A generalized Hermite normal form $\C\in \Zx^{n\times s}$ given in \bref{ghf}.}\\
  \Outp{$\emptyset$, if $L = (\C)_{\Zx}$ is  $\Z[x]$-saturated;
      otherwise, a finite set $\{\h_1,\ldots,\h_r\}\subset\Zxn$
    such that $\h_i\in \sat_{\Z[x]}( L)\setminus L$, $i=1,\ldots,r$.}\medskip

  \noindent
  1. Let $S=${\bf{ZFactor}}$(\C)$. If $S\ne\emptyset$ return $S$. \\
  2. For any prime factor $p(x)\in\Zx\setminus\Z$ of $\prod_{i=1}^t c_{r_i,1}$,
  execute steps 2.1-2.3,
  where $c_{r_i,1}$ are from \bref{ghf}.\\
   \SPC 2.1. Set $M=[\c_{r_1,1},\ldots,\c_{r_t,1}]\in \Z[x]^{n\times t}$,
   where $\c_{r_i,1}$ can be found in \bref{ghf}.\\
   \SPC 2.2. Compute a finite basis $B=\{\b_1,\ldots,\b_l\}$ of $\Syz(M)=\{X\in\Q[x]^t \,|\,MX=0\}$\\
    \SPC\SPC as a $K$-vector space in $K^t$, where $K=\Q[x]/(p(x))$.\\
   \SPC 2.3. If $B\ne\emptyset$,\\
   \SPC\SPC 2.3.1.   For each $\b_i$, let $M\b_i= p(x)
   \frac{\g_i}{m_i}$,
     where $\g_i\in\Zxn$ and $m_i\in\Z$.\\
   \SPC\SPC 2.3.2. Return $\{\g_1,\ldots,\g_l\}$ \\
  3. Return $\emptyset$.
\smallskip
\end{algorithm}

The {\em $\Zx$-saturation} of a $\Zx$-lattice $L$ is defined to be
$$\sat_{\Zx}(L)=\{\f\in
\Z[x]^{n}\, |\, \exists p \in \Zx\backslash\{0\} \st p\f\in L\}.$$
The following algorithm compute $\sat_{\Zx}(L)$.
\begin{algorithm}[H]\label{alg-satZX}
  \caption{\bf --- SatZX$(\bu_1,\ldots,\bu_m)$} \smallskip
  \Inp{A finite set $U=\{\bu_1,\ldots,\bu_m\}\subset\Zxn$.}\\
  \Outp{A set of generators of $\sat_{\Z[x]}(L)$, where $L=(U)_{\Zx}$.}\medskip

  \noindent
  1. Compute a generalized Hermite normal form $\gb$ of $U$ \cite{ghnf}.\\
  2. Set $S=${\bf ZXFactor}$(\gb)$.\\
  3. If $S=\emptyset$, return $\gb$; otherwise set $U=\gb \cup S$ and go to step 1.
\smallskip
\end{algorithm}

\begin{example}
Let \[ \C
 = \left[  \begin{array}{llllll}
    x              & x^2+1     \\
    2x^2+1            & 0    \\
    0                 & 4x^2+2 \\
\end{array} \right]. \]
Apply Algorithm~{\bf{ZXFactor}} to $\C$. In step, 1, $S=\emptyset$
and $\C$ is $\Z$-saturated.
In step 2, the only irreducible factor of $\prod_{i=1}^t
c_{r_i,1}\in\Zx$ is $p(x)=2x^2+1$.
In step 2.1, $M=\C$ and in step 2.2, $B=\{[-1,2x]^\tau\}$. In step
2.3.1, $M\cdot [-1,2x]^\tau =  2x\c_{2,1}-\c_{1,1} =
p(x)[x,-1,4x]^\tau = 0~\mod ~p(x)$ and $\{[x,-1,4x]^\tau\}$ is
returned.

In Algorithm {\bf{SatZX}}, $\h=[x,-1,4x]^\tau$ is added into $ \C $
and the generalized Hermite normal form of $\C\cup\{\h\}$ is
 \[ \C_1
 = \left[  \begin{array}{llllll}
    x         & 1    \\
    2x^2+1    & x   \\
    0          & 2 \\
\end{array} \right]. \]
Apply Algorithm~{\bf{ZXFactor}} to $\C_1$,  one can check that
$\C_1$ is $\Z[x]$-saturated.
\end{example}

In the rest of this section, we will prove the correctness of the
algorithm.
Similar to the definition of $\sat_{\Z[x]}(L)$, we can
define $\sat_{\Q[x]}(L_{\Q[x]})$. $L_{\Q[x]}$ is called {\em $\Q[x]$-saturated}
if $\sat_{\Q[x]}(L_{\Q[x]})=L_{\Q[x]}$. The following lemma
gives a criterion for whether $L$ is $\Zx$-saturated.
\begin{lemma}\label{lm-satzx1}
A $\Zx$-lattice $L$ is $\Zx$-saturated if and only if $\sat_\Z(L) =
L$ and $\sat_{\Q[x]} (L_{\Q[x]}) = L_{\Q[x]}$.
\end{lemma}
\proof
%
$``\Rightarrow"$ If $L=(\{\bu_1,\ldots, \bu_m\})_{\Zx}$ is $\Zx$-saturated, then
$\sat_\Z(L) = L$. If $\sat_{\Q[x]} (L_{\Q[x]}) \ne L_{\Q[x]}$, then there
exists an $h(x)\in\Q[x]$ and a $\g\in\Q[x]^n$, such that $h(x)\g\in
L_{\Q[x]}$ but $\g\not\in L_{\Q[x]}$. From $h(x)\g\in L_{\Q[x]}$, we have $h(x)\g =
\sum_{i=1}^s q_i(x)\bu_i$ where $q_i(x)\in\Q[x]$. Clearing the
denominators of the above equation, there exist $m_1, m_2\in \Z$
such that $m_1h(x)\in\Zx$, $m_2\g\in\Zxn$, and $m_1h(x)\cdot
m_2\g\in L$. Since $L$ is $\Zx$-saturated, $m_2\g\in L$, which
contradicts to $\g\not\in L_{\Q[x]}$.

$``\Leftarrow"$ For any $h(x)\in\Zx$ and $\g\in\Zxn$,  if $h(x)\g\in
L$, we have $h(x)\g\in L_{\Q[x]}$, and hence $\g\in L_{\Q[x]}$ since
$\sat_{\Q[x]} (L_{\Q[x]}) = L_{\Q[x]}$. From $\g\in L_{\Q[x]}$, there exists an
$m\in \Z$ such that $m\g\in L$ which implies $\g\in L$ since $L$ is
$\Z$-saturated. \qedd

%

In the following two lemmas,  $\C$ is the generalized Hermite normal form given in \bref{ghf}.

\begin{lemma}\label{lm-satzx2}
$(\C)_{\Q[x]}=(\c_{r_1,1},\ldots,$ $\c_{r_t,1})_{\Q[x]}$.
\end{lemma}
\proof
%
We will prove $(\C)_{\Q[x]}=(\c_{r_1,1},\ldots,$ $\c_{r_t,1})_{\Q[x]}$ by induction.
By 3) of Definition \ref{def-ghf}, $S(\c_{r_1,1},\c_{r_1,2}) = x^u
\c_{r_1,1} - a\c_{r_1,2}$ ($u\in\N$ and $a\in\Z$) can be reduced to
zero by $\c_{r_1,1}$, which means $\c_{r_1,2} = q(x)\c_{r_1,1}$
where $q(x)\in \Q[x]$. Hence, $(\c_{r_1,1},\c_{r_1,2})_{\Q[x]}=
(\c_{r_1,1})_{\Q[x]}$ as $\Q[x]$-modules.
Suppose for $k<l_1$, $(\c_{r_1,1},\ldots,\c_{r_1,k})_{\Q[x]}= (\c_{r_1,1})_{\Q[x]}$
as $\Q[x]$-modules. We will show that
$(\c_{r_1,1},\ldots,\c_{r_1,k+1})_{\Q[x]}= (\c_{r_1,1})_{\Q[x]}$ as $\Q[x]$-modules.
Indeed, by 3) of Definition \ref{def-ghf},
$S(\c_{r_1,1},\c_{r_1,k+1}) = x^v \c_{r_1,1} - b\c_{r_1,k+1}$
($v\in\N$ and $b\in\Z$) can be reduced to zero by
$\c_{r_1,1},\ldots,\c_{r_1,k}$ and hence, $\c_{r_1,k+1} \in
(\c_{r_1,1})_{\Q[x]}$. Then we have
$(\c_{r_1,1},\ldots,\c_{r_1,l_1})_{\Q[x]}= (\c_{r_1,1})_{\Q[x]}$.
For the rest of the polynomials in $\C$, the proof is similar. \qedd

The following lemma gives a criterion for a $\Q[x]$-module to be
$\Q[x]$-saturated.
\begin{lemma}\label{lm-satzx3}
Let  $L=(\C)_{\Q[x]}$. Then $\sat_{\Q[x]}(L)=L$ if and only if
$\C_1=\{\c_{r_1,1},\ldots,\c_{r_t,1}\}$ is linear independent over
$\K_{p(x)}=\Q[x]/(p(x))$ for any irreducible polynomial
$p(x)\in\Z[x]$.
\end{lemma}
\proof  $``\Rightarrow"$ Assume the contrary, that is, $\C_1$ are
linear dependent over $\K_{p(x)}$ for some $p(x)$. Then there exist
$g_i\in\Q[x]$  not all zero in $\K_{p(x)}$, such that $\sum_{i=1}^t
g_i\c_{r_i,1} = 0$ in $\K_{p(x)}^n$ and hence $\sum_{i=1}^t g_i
\c_{r_i,1} = p(x)\g$ in $\Q[x]^n$.
Since $\C_1$ is in upper triangular form and is clearly linear independent
in $\Q[x]^n$, we have $\g\ne{\bf{0}}$.
Since $\sat_{\Q[x]}(L)=L$, we have $\g\in L$.
Then, there exist $f_i\in\Q[x]$ such that $\g = \sum_{i=1}^t
f_i\c_{r_i,1}$. Hence $\sum_{i=1}^t(g_i-pf_i)\c_{r_i,1} = 0$ in
$\Q[x]^n$. Since $\C_1$ is linear independent in $\Q[x]^n$, $g_i =
pf_i$ and hence $g_i=0$ in $\K_{p(x)}$, a contradiction.

$``\Leftarrow"$ Assume the contrary, that is, there exists a
$\g\in\Q[x]^n$, such that $\g\not\in L$ and $p(x)\g \in L$
for an irreducible polynomial $p(x)\in\Z[x]$. Then, by Lemma
\ref{lm-satzx2} we have $p\g = \sum_{i=1}^t f_i\c_{r_i,1}$, where
$f_i\in\Q[x]$. $p$ cannot be a factor of all $f_i$. Otherwise, $\g =
\sum_{i=1}^t \frac{f_i}{p}\c_{r_i,1}\in L$. Then some of $f_i$
is not zero in $\K_{p(x)}$, which means $\sum_{i=1}^t
f_i\c_{r_i,1}=0$ is a nontrivial linear relation among $\C_1$ over
$\K_{p(x)}$, a contradiction. \qedd

From the $``\Rightarrow"$  part of the above proof, we have
\begin{cor}\label{cor-satzx1}
Let $\C$ be the generalized Hermite normal form given in \bref{ghf}
and $\sum_{i=1}^t f_i \c_{r_i,1}$ $=0$ a nontrivial linear relation
among $\c_{r_i,1}$ in $(\Q[x]/(p(x)))^n$, where $p(x)$ is an
irreducible polynomial in $\Zx$ and  $f_i\in \Q[x]$. Then, in
$\Q[x]^n$, $\sum_{i=1}^r f_i \c_{r_i,1}=p(x)\g$ and $\g\not\in (\C)_{\Q[x]}$.
\end{cor}

\begin{theorem}
Algorithms~{\bf SatZX} and {\bf ZXFactor} are correct.
\end{theorem}
\proof In Step 3 of  Algorithm~{\bf SatZX},
if $(\gb)_{\Zx}$ is not $\Zx$-saturated, then  $S\ne\emptyset$ and  $(\gb)_{\Zx}\nsubseteq(\gb\cup S)_{\Zx}\subset\sat_{\Zx}((\gb)_{\Zx})$.
Since $\Zxn$ is Notherian, the algorithm will terminate and
outputs $\sat_{\Zx}(L)$.
Thus, it suffices to prove the correctness of Algorithm {\bf ZXFactor}.

In step 1 of Algorithm {\bf ZXFactor}, if $S\ne\emptyset$,
then from properties of Algorithm {\bf ZFactor}, $S\subset\sat_{\Z}((\C)_{\Zx})\setminus(\C)_{\Zx} \subset\sat_{\Z[x]}((\C)_{\Zx})\setminus(\C)_{\Zx}$.
The algorithm is correct.
In step 2, we claim that $L$ is $\Q[x]$-saturated if and only if
$B=\emptyset$ and if $B\ne\emptyset$ then $\g_i$ in step 2.3.1 is
not in $L$. In step 3, $L$ is both $\Z$- and $\Q[x]$-saturated. By
Lemma \ref{lm-satzx1}, $L$ is $\Z[x]$-saturated and the algorithm is
correct. So, it suffices to prove the claim about step 2.

Let $L=(\C)_{\Zx}$.
In Step 2, $L$ is already $\Z$-saturated.
Then by Lemma \ref{lm-satzx1}, $L$ is $\Zx$-saturated if and only if
$(\C)_{\Q[x]}$ is $\Q[x]$-saturated.
By Lemma~\ref{lm-satzx3}, to check whether
$(\C)_{\Q[x]}$ is $\Q[x]$-saturated, we  need only to check whether for
any irreducible polynomial $p(x)\in\Zx$,
$\C_{1}=\{\c_{r_1,1},\ldots,\c_{r_t,1}\}$  is linear independent
over $\K_{p(x)} = \Q[x]/(p(x))$.
If  $p(x)$ is not a prime factor of  $\prod_{i=1}^t c_{r_i,1}$, then
the leading monomials of $\c_{r_i,1}, i=1,\ldots,t$ are nonzero
and $\C_1$ is in upper triangular form.
As a consequence, $\C_1$ must be linear independent over $\K_{p(x)}$.
Then, in order to check whether $L$ is $\Q[x]$-saturated,
it suffices to consider prime factors of
$\prod_{i=1}^t c_{r_i,1}$ in step 2 of the algorithm.
In step 2.3, it is clear that if $B=\emptyset$ then $\C_1$ is
linear independent over $\K_{p(x)}$.
For $\b_i\in B$, since $M\b_i=0$ over $\K_{p(x)}$, $M\b_i = p(x)
\h_i$ where $\h_i\in\Q[x]^t$. Hence $\h_i = \frac{\g_i}{m_i}$ for
$\g_i\in\Zx^t$ and $m_i\in\Z$.
By Corollary \ref{cor-satzx1}, $\g_i\not\in L$.
Therefore, step 2 returns a set of nontrivial factors of $L$ if $L$
is not $\Z[x]$-saturated. The claim about step 2 is proved. \qedd

\section{Conclusion}
\label{sec-conc}
In this paper, we initiate the study of  toric $\sigma$-varieties.
A toric $\sigma$-variety is defined as the Cohn closure of the values
of a set of Laurent $\sigma$-monomials. Three characterizing properties
of toric $\sigma$-varieties are proved in terms of its coordinate
ring, its defining ideals, and group actions.
In particular, a
$\sigma$-variety is toric if and only if its defining ideal is
a toric $\sigma$-ideal, meaning a binomial $\sigma$-ideal whose
support lattice is $\Zx$-saturated.
Algorithms are given to decide whether the binomial $\sigma$-ideal
$\I_L$ with support lattice $L$ is toric.

We establish connections between toric $\sigma$-varieties and affine $\N[x]$-semimodules.
We show that
the category of toric $\sigma$-varieties with toric morphisms is antiequivalent to the category of affine $\N[x]$-semimodules with $\N[x]$-semimodule morphisms.
We also show that there is a one-to-one correspondence between irreducible $T$-invariant subvarieties of a toric $\sigma$-variety $X$ and faces of the corresponding affine $\N[x]$-semimodule, where $T$ is the $\sigma$-torus of $X$. Besides, there is also a one-to-one correspondence between $T$-orbits of the toric $\sigma$-variety $X$ and faces of the corresponding affine $\N[x]$-semimodule $S$, when $S$ is face-saturated.

\end{document}